\documentclass[twocolumn,aps, pra, nofootinbib,superscriptaddress,showkeys,showpacs,longbibliography]{revtex4-1}
\usepackage{amsmath,amsfonts,amssymb,amstext,amscd,amsthm,dsfont}
\usepackage{color}
\usepackage[colorlinks, citecolor=blue]{hyperref}
\usepackage{amsfonts}
\usepackage[font=small,labelfont=bf,justification=justified]{caption}
\usepackage{braket}
\usepackage{graphicx}
\usepackage{subcaption}
\usepackage{times}
\usepackage{color}
\usepackage{soul,xcolor}
\usepackage{bm}
\usepackage{gensymb}
\usepackage[normalem]{ulem}
\usepackage{ragged2e}

\begin{document}

\title{Accelerated quantum walk, two-particle entanglement generation and localization}

\author{Shivani Singh}
\affiliation{The Institute of Mathematical Sciences, C. I. T, Campus, Taramani, Chennai, 600113, India} 
\affiliation{Homi Bhabha National Institute, Training School Complex, Anushakti Nagar, Mumbai 400094, India}
\email{shivanis@imsc.res.in}
\author{Radhakrishnan Balu}
\affiliation{U.S. Army Research Laboratory, Computational and Information Sciences Directorate, Adelphi, Maryland 20783, USA}
\affiliation{Computer Science and Electrical Engineering, University of Maryland Baltimore County, 1000 Hilltop Circle, Baltimore, Maryland 21250, USA}
\author{Raymond Laflamme}
\affiliation{Institute for Quantum Computing and Department of Physics and Astronomy, University of Waterloo, Waterloo N2L 3G1, Ontario, Canada}
\affiliation{Perimeter Institute for Theoretical Physics, Waterloo, N2L 2Y5, Ontario, Canada}
\author{C. M. Chandrashekar}
\email{chandru@imsc.res.in}
\affiliation{The Institute of Mathematical Sciences, C. I. T, Campus, Taramani, Chennai, 600113, India}
\affiliation{Homi Bhabha National Institute, Training School Complex, Anushakti Nagar, Mumbai 400094, India}


\begin{abstract}
We present a scheme to describe the dynamics of accelerating discrete-time quantum walk for one- and two-particle in position space. We show the effect of acceleration in enhancing the entanglement between the particle and position space in one-particle quantum walk and in generation of entanglement between the two unentangled particle in two-particle quantum walk. By introducing the disorder in the form of phase operator we study the transition from localization to delocalization as a function of acceleration. These inter-winding connection between acceleration, entanglement generation and localization along with well established connection of quantum walks with Dirac equation can be used to probe further in the direction of understanding the connection between acceleration, mass and entanglement in relativistic quantum mechanics and quantum field theory. Expansion of operational tools for quantum simulations and for modelling quantum dynamics of accelerated particle using quantum walks is an other direction where these results can play an important role. 
\end{abstract}

\maketitle
\vskip -0.7in
\noindent

\section{\label{sec1} Introduction}

Quantum walks \cite{GVR, RPF, ADZ, DAM} have played an important role in development of various quantum information processing and computation protocols \cite{JK, ESV,ANAV, IKS}.  It is also among the most promising candidate to model controlled quantum dynamics and understand them from quantum information perspective.  One can see quantum walks as a very promising protocol for quantum simulations in parallel to that of the classical random walk which has played an important role in classical simulations over decades now. Demonstration of realization of quantum walks in a variety of physical systems, such as NMR\,\cite{RLBL}, trapped ions\,\cite{SMS09, ZKG10}, integrated photonics\,\cite{AS,BFL,PLM10,PLP}, and bulk optics\,\cite{KFC09} in recent years guarantee a promising role it can play in broad field of future quantum technologies. 

Quantum walks evolves particle in extended position space making effective use of quantum phenomena like superposition and interference. This results in quadratically faster spreading of  wavepacket in position when compared to classical random walk \cite{JK, ESV, YKE}.   Analogues to discrete version of classical random walk, discrete-time quantum walk spreads in position Hilbert space with probability amplitude assigned using a quantum coin operation on the particle (walker) Hilbert space. The general unitary coin operation can have a three independent parameters \cite{CSL08} which can help in providing more control over the dynamics.  These simple dynamics has proved to be a strong algorithmic techniques in modelling the dynamics of for example,  photosynthesis \cite{MRL}, diffusion in quantum system \cite{GF},  localization (Anderson localization and Weak localization) \cite{AVW, SL, WL, AE, AJ, GB,CMC,OKA,JM,NK,Crespi,SK,AE}, topological phases \cite{STL,SRF,AO, COT}, Dirac equations and associated dynamics\,\cite{Str06, CBS10, MC16, AP14, Per16, PBL18, bia94, Cha13}. 

In this work we will introduce acceleration in the discrete-time quantum walk dynamics by introducing a time-dependent parameter into the quantum coin operations. Changing the acceleration parameter changes the way probability amplitude spreads in the position space and we observe some interesting features like enhancement in the quantity of entanglement between the particle and position space as a function of acceleration in one-dimensional single-particle quantum walk. We also study the two interacting particle discrete-time quantum walk and present the situations where acceleration plays a role in entangling the two initially unentangled particles. These results are in general valid for any time dependent function of coin parameter which introduces acceleration into the dynamics. The interacting coin operator we have used in our two particle dynamics without acceleration parameter in its Hamiltonian form is equivalent to the Hamiltonian of Ising model which is a special case of generalised Heisenberg model \cite{LSM, KD}.  

Disorder in any of the parameter describing the single-particle discrete-time quantum walk dynamics has been studied in  past\,\cite{VFQ, SD}. The disordered discrete-time quantum walk leading to localization of the particle in position space resulting in Anderson localization (strong localization) and weak localization \cite{CM, SC, RAS, OKA, JM, YKE, NK} has also been extensively reported. Here, using the quantum coin operator followed by a phase operator we will introduce both, acceleration and disorder, respectively into the quantum walk dynamics. This enabled us to study the interplay between the acceleration and localization of the single-particle quantum walker.  We further extend our study to two-particle walker and show the role of localization in enhancing and preserving the entanglement between the two initially unentangled particle for larger number of steps of quantum walk. Anderson localization of entangled particle has already been implemented in experiment \cite{Crespi}. Here we are presenting a scheme to implement two-particle quantum walk where the two particles are initially unentangled but after few steps of walk they entangle and how disorder in such walk effects the localization. These inter-winding connection between acceleration, entanglement generation and localization along with well established connection of quantum walks with massive Dirac equation \cite{MC16} can be used to probe further in the direction of understanding the connection of acceleration, mass and entanglement in relativistic quantum mechanics and quantum field theory. These results will further guide us in developing an operational tools for quantum simulations and for modelling the dynamics of range of accelerated quantum particle using quantum walks. Our work presents both, analytical study followed by analysis of the dynamics  from the dispersion relations and numerical results complements where analytical results were not attainable. 

\begin{figure}[h!]
\centering
\includegraphics[width = \linewidth]{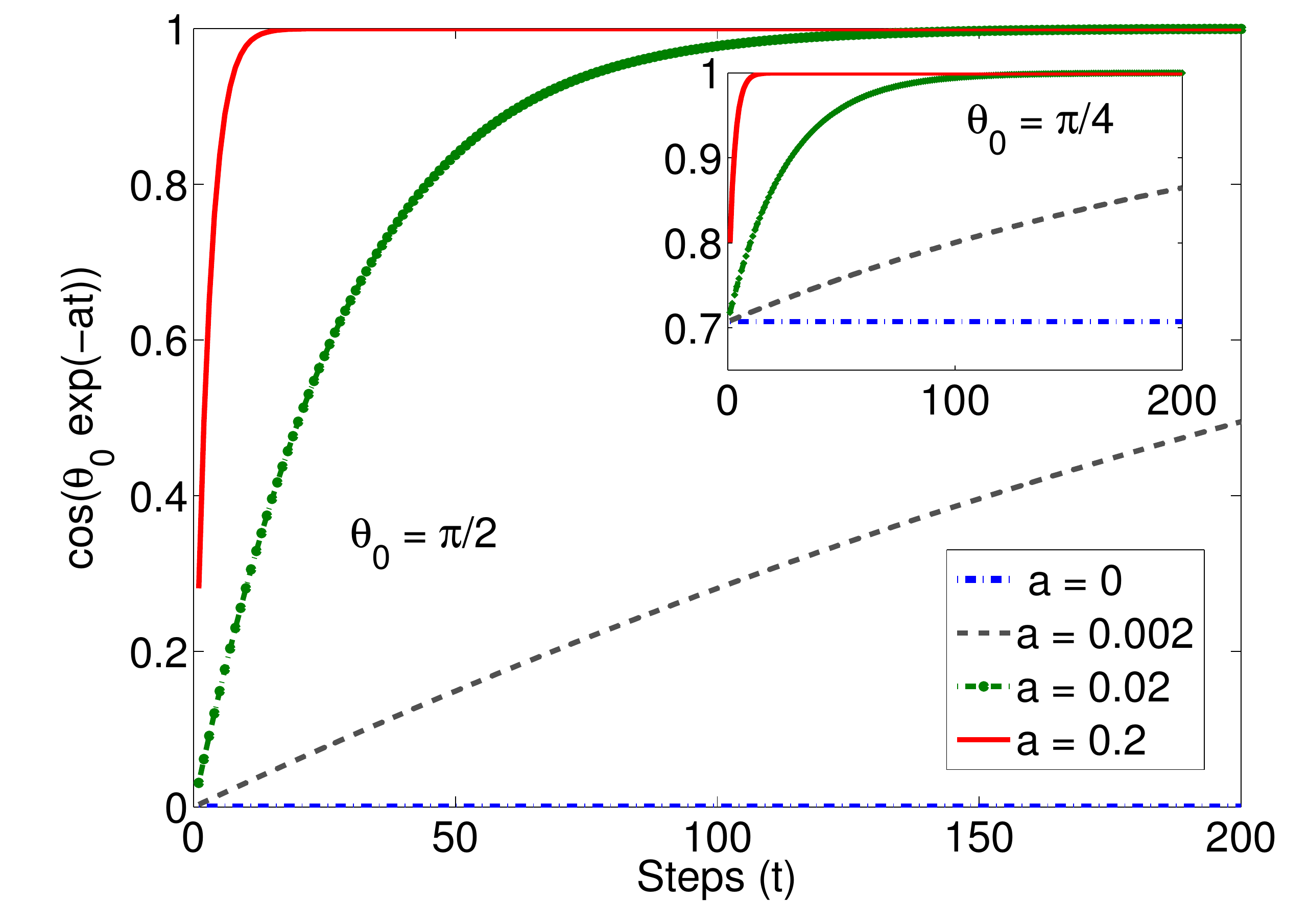}
\caption{Effect of parameter $a$ which induces acceleration on the quantum walker as function of $t$ is shown by plotting $\cos(\theta_0 e^{-at})$. For higher value of $a$ we can see a value reaching unity at earlier time.  This gives us a range of values $a$ to look into to understand the effect of acceleration on the quantum walk dynamics.}
\label{1D_effctacce}
\end{figure}
In section \ref{sec2} we introduce accelerated discrete-time quantum walk in one dimension and two dimension for one-particle and two-particle quantum walk, respectively. We have described the single-particle accelerated quantum walk and have shown how the acceleration changes the probability distribution and its effect on entanglement between the single particle and position space. Dynamics of the two-particle accelerated walk in two-dimensional and the interaction between the particles and its confinement to one spatial dimension due to bosonic/fermionic nature is also introduced in the same section. In section \ref{sec3}, we have calculated the dispersion relation and the transfer matrix for accelerated discrete-time quantum walks for single- and two-particles. This shows how the probability amplitude changes from one position to an another for a given time step and also the dependence of probability amplitude on the parameter of the evolution operator. Then we have introduced temporal and spatial disorder in the one of the parameter of evolution operator (coin operator) and acceleration in the other coin parameter to study weak and strong (Anderson) localization in the accelerated discrete-time quantum walk. For accelerated quantum walk, probability amplitude spreads faster over larger space as we increase acceleration. In both, section \ref{sec2} and section \ref{sec3} we have studied the entanglement between the two-particles after tracing out the position space and entanglement between the particle and position space in two-particle accelerated discrete-time quantum walk. This shows that even when we start with an unentangled initial state, after few steps of quantum walk we find the particles to be entangled but as the time increases the entanglement between the particles decays for accelerated walk. The decay time is lower as we increase the acceleration. Similarly, for disordered two-particle accelerated quantum walk, the decay time is lower as we decrease the acceleration. Entanglement between the particle and position space shows similar characteristic as single-particle accelerated walk in one dimension.  In section \ref{sec4} we amalgamate our observations and conclude.
 
\begin{figure}[h!]
\centering
\includegraphics[width = \linewidth]{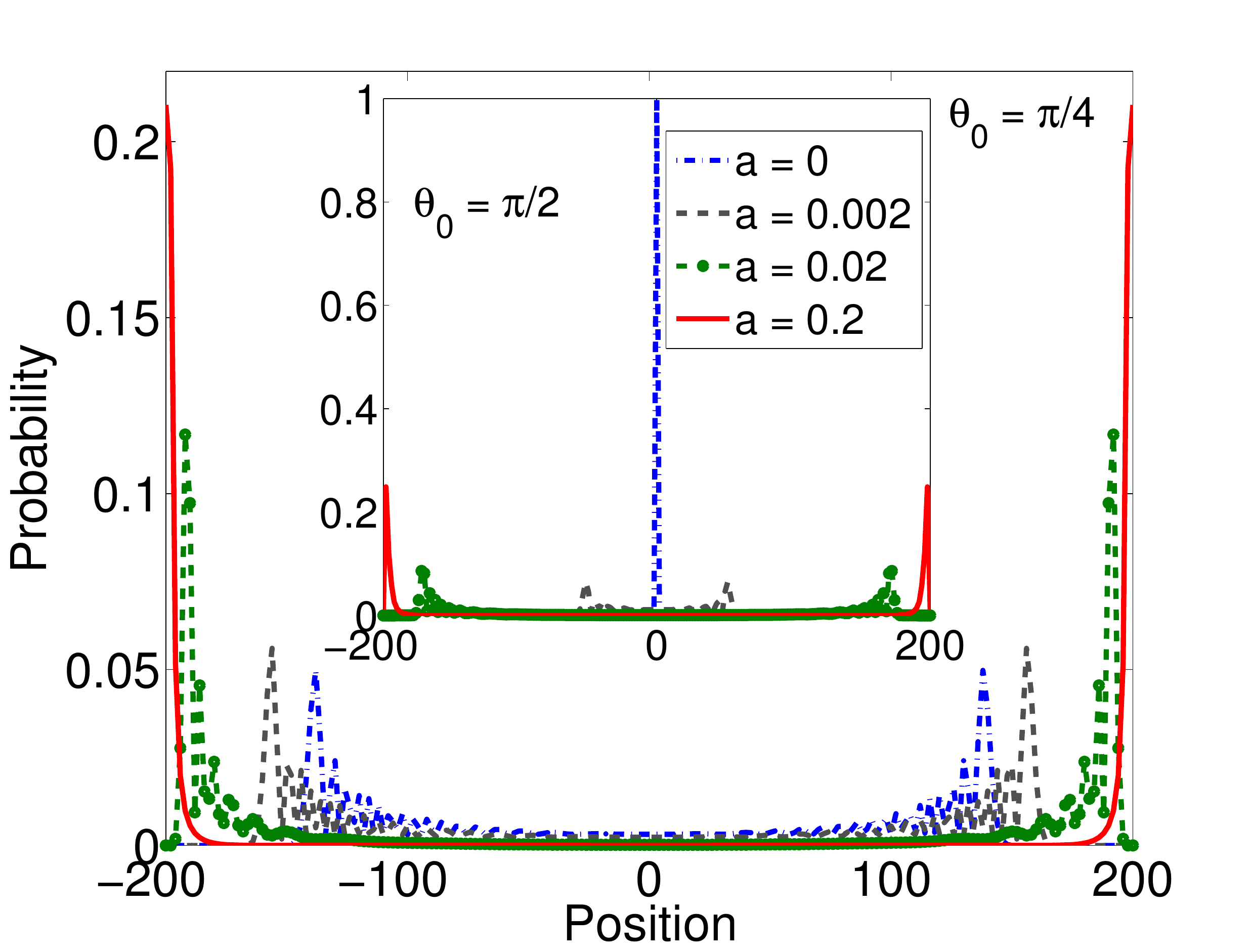}
\caption{Probability distribution of an one-dimensional accelerated quantum walk with $\theta(a, t) = \theta_0 e^{-at}$ for different value of $a$ after $t = 200$ steps of walk. Parameter $a=0$ corresponds to evolution with constant velocity, an homogeneous evolution and the minimum spread is set by the value of $\theta_0$. With increase in $a$, spread in position space increases. The initial state is $\ket{\Psi_{in}} = \frac{1}{\sqrt 2}(\ket{\uparrow} + \ket{\downarrow}) \otimes \ket{x = 0}$. Probability distribution is shown for $\theta_0 = \pi/4$ and $\theta_0 = \pi/2$ (inset). The spread is minimum for $a=0$ and when $\theta_0 = \pi/2$, it is localized at the origin $x=0$. }
\label{1D_prob}
\end{figure}

\section{\label{sec2} Accelerated Discrete-time Quantum walk}

\subsection{\label{subsec1} Single-particle}

Evolution of a single-particle discrete-time quantum walk is defined on a Hilbert space $\mathcal{H}$ which is a composition of a particle (coin) Hilbert space and the position Hilbert space, $\mathcal{H} = \mathcal{H}_{c} \otimes  \mathcal{H}_p$.  The coin Hilbert space $\mathcal{H}_{c}$ represents  the addressable basis state of the particle which is subjected to the quantum walk and the position Hilbert space $\mathcal{H}_p$ represents the position basis states on which the particle can move in superposition of position space.  For a two level particle as quantum walker in one-dimensional position space, the $\mathcal{H}_{c}$ is spanned by the basis states $\left\lbrace \ket{\uparrow}, \ket{\downarrow} \right\rbrace$ and $\mathcal{H}_p$ is spanned by the basis state $\left\lbrace \ket{\it{x}} \right\rbrace$ where $\it{x} \in \mathbb{Z}$.  The initial state of particle on which we define the quantum walk can be written in the form,
\begin{equation}
|\Psi_{in}\rangle  = (\alpha |\uparrow \rangle + \beta |\downarrow \rangle ) \otimes |x = 0\rangle. 
\end{equation}
The evolution operator which defines each step of the standard or homogeneous discrete-time quantum walk is composed of a quantum coin operator,
\begin{equation} \label{coin_1D}
C({\theta_0}) = \begin{pmatrix}
~~~\cos(\theta_0) & -i\sin(\theta_0)\\
-i\sin(\theta_0) & ~~~\cos(\theta_0)
\end{pmatrix}  \otimes \sum_{x}\ket{x}\bra{x}
\end{equation}
followed by a position shift operator 
\begin{equation}
S_x = \sum_x \Big[ \ket{\uparrow}\bra{\uparrow} \otimes  \ket{\textit{x}-1}\bra{\textit{x}} + \ket{\downarrow}\bra{\downarrow} \otimes  \ket{\textit{x}+1}\bra{\textit{x} } \Big].
\end{equation} 
After $t$ steps of quantum walk, the state will be in the form,
\begin{equation} \label{Gen_Psi}
|\Psi_t \rangle = \Big (S_x C(\theta_0)\Big )^t |\Psi_{in}\rangle = \sum_{x} \Big ( \mathcal{A}_{x, t}|\uparrow  \rangle + \mathcal{B}_{x, t} |\downarrow \rangle \Big ) \otimes |x \rangle
\end{equation}
where $\mathcal{A}_{x, t}$ and  $\mathcal{B}_{x, t}$ are the probability amplitudes of the states $|\uparrow\rangle$ and $|\downarrow\rangle$ at position $x$ and time $t$. Amplitudes $\mathcal{A}_{x, t}$ and $\mathcal{B}_{x, t}$ as a function of amplitude at its neighbouring position at previous time $(t-1)$ can be written as,
\begin{align} 
\mathcal{A}_{x, t} & =  \cos(\theta_0) \mathcal{A}_{x+1, t-1}  - i  \sin(\theta_0) \mathcal{B}_{x+1, t-1} \label{Gen_PsiA1}\\
\mathcal{B}_{x, t} &= -i \sin(\theta_0) \mathcal{A}_{x-1, t-1} + \cos(\theta_0) \mathcal{B}_{x-1, t-1}. \label{Gen_PsiA2}  
\end{align}
The probability distribution in position space at time $t$, $P_{x, t} = \lvert  \mathcal{A}_{x, t} \rvert ^2 + \lvert  \mathcal{B}_{x, t} \rvert^2$ will spread quadratically faster in bimodal form when compared to Gaussian spread of the classical random walk~\cite{ANAV}.   However, deviation from the bimodal distribution can be obtained in multiple ways. For example, by increasing the dimension of $\mathcal{H}_{c}$  to three and defining the corresponding evolution operator, an additional localized mode can be obtained~\cite{IKS}. By introducing coin operation with randomly picked coin parameter for different position (spatial disorder) or for different time (temporal disorder), a strongly localized or a weakly localized probability distribution can be seen~\cite{CMC, RAS, OKA, JM, YKE, NK, SK}. Thus, with the controllable quantum walk evolution parameter one can demonstrate a good control over the dynamics and probability distribution of the quantum walker.

In Fourier mode, wave like solution form the probability amplitude $\mathcal{A}_{x,t}$ and $\mathcal{B}_{x,t}$ can in general be written as $\psi_{x,t} = e^{i(-\omega_1 t + \kappa_1 x)} \psi(\kappa_1)$ where $\omega_1$ is the wave frequency and $\kappa_1$ is the wave number and $(\psi_{x,t}^{\uparrow}; \psi_{x,t}^{\downarrow}) \equiv (\mathcal{A}_{x,t}; \mathcal{B}_{x,t})$. Substituting the Fourier form of the probability amplitude in Eq. \eqref{Gen_PsiA1} and \eqref{Gen_PsiA2} and solving for a relation between $\kappa_1$ and $\omega_1$ we get the dispersion relation of the form,
\begin{align} \label{eqDRS}
\cos(\omega_1) = \cos(\theta_0)\cos(\kappa_1)
\end{align}
which implies that the group velocity for the probability amplitude is given by,
\begin{align} \label{eqGV}
v_g = \frac{\cos(\theta_0) \sin(\kappa_1)}{\sqrt{1 - (\cos(\theta_0)\cos(\kappa_1))^2}}.
\end{align}
Maximizing the group velocity with respect to $\kappa_1$, we find that $v_g$ is maximum for $\kappa_1 = \pi/2$, 
\begin{align} \label{EqMaxVel}
v_g = \cos(\theta_0).
\end{align}

\begin{figure}[h!]
\centering
\includegraphics[width = \linewidth]{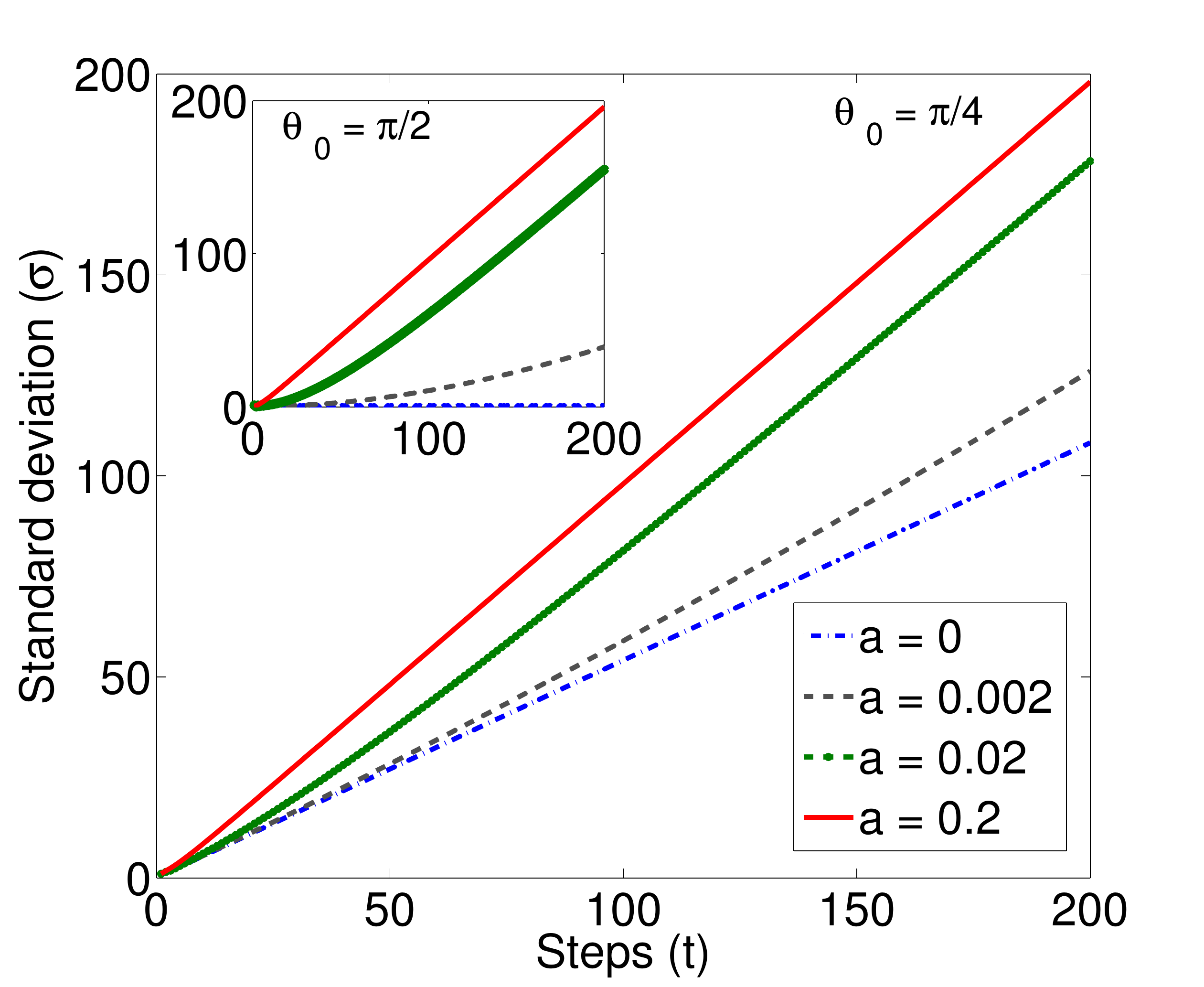}
\caption{Standard deviation of an one-dimensional accelerated quantum walk with $\theta(a, t) = \theta_0 e^{-at}$ for different value of $a$ as function of $t$ (steps). The initial state is $\ket{\Psi_{in}} = \frac{1}{\sqrt 2}(\ket{\uparrow} + \ket{\downarrow}) \otimes \ket{x = 0}$. Standard deviation is  shown for two values of $\theta_0 = \pi/4$ and $\pi/2$ (inset). With increase in $a$ and $t$ increase in standard deviation is seen and the increase is more pronounced then $\theta_0 = \pi/2$.}
\label{1D_sdsteps}
\end{figure}
\begin{figure}[h!]
\centering
\includegraphics[width = \linewidth]{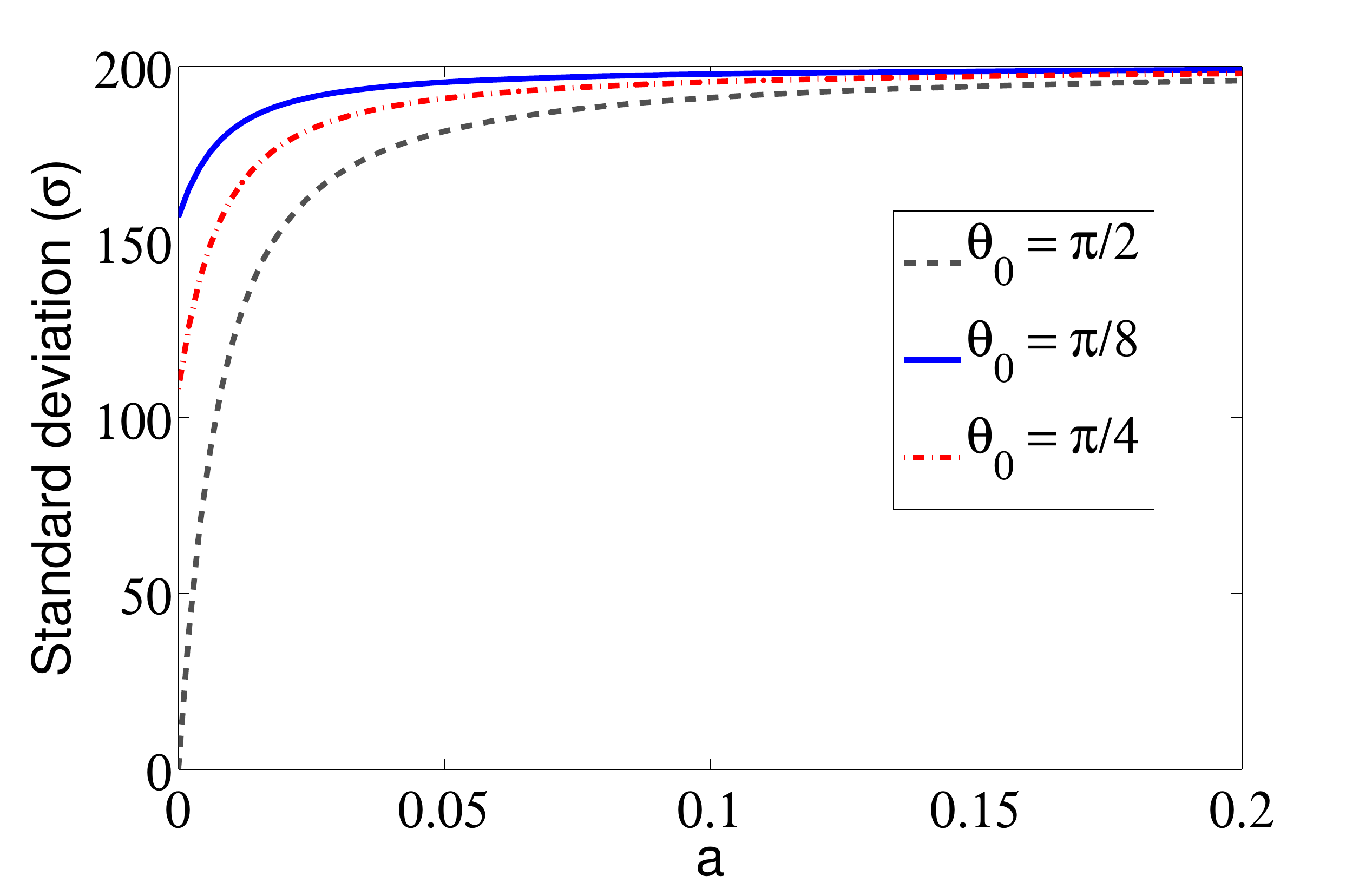}
\caption{Standard deviation of an one-dimensional accelerated quantum walk with $\theta(a, t) = \theta_0 e^{-at}$ for different value of $\theta_0$ as function of $a$ when $t = 200$. The initial state is $\ket{\Psi_{in}} = \frac{1}{\sqrt 2}(\ket{\uparrow} + \ket{\downarrow}) \otimes \ket{x = 0}$. With increase in $a$ an increase in $\sigma$ is seen until it reaches a maximum value ($\sigma = t$).}
\label{1D_sdacce}
\end{figure}
\noindent 
When $\theta_0 = \pi/2$ in the coin operation of the form given by Eq.\,(\ref{coin_1D}),  the amplitude of quantum walker will remain localized between position $x=0$ and $x=\pm1$ for all time $t$. When $\theta_0 = 0$, the two states of the walker will move away from each other and particle will only be seen at  the positions $x = \pm t$ with non-zero probability.  For both these values of $\theta$ we will not see any interference effect in the dynamics and they define the bounds on the spreading of the walk. For any value of $\theta$ between these values, interference effects plays an important role and the spread of the probability distribution of the walker after $t$ steps of walk will be bounded between $\pm t \cos(\theta_0)$ ~\cite{ANAV, CSL08} and this is also evident from Eq.\,\eqref{EqMaxVel} where for given value of $\theta_0$, the maximum value of the group velocity is $\cos(\theta_0)$.  Since $\theta$ in the coin operation plays an important role in defining the dynamics of the walker, by making $\theta$ a time dependent parameter we can construct a quantum walk evolution which results in upward change of group velocity for each increasing instant of time. This upward change in group velocity will introduce instantaneous acceleration to the quantum walk dynamics. Therefore, acceleration to quantum walk dynamics can be introduced any time dependent function of coin parameter which contributes change in group velocity with time. Here, we will choose a time dependent coin parameter of the form
\begin{equation}
 \theta(a, t) = \theta_0 e^{-at}
 \end{equation}
 in place of $\theta_0$ in the standard quantum coin evolution operator, Eq.\,(\ref{coin_1D}). The parameter $a$ in the coin operation induces acceleration to the dynamics where the minimum spread will be bounded by the value of $\theta_0$ and maximum spread will be achieved with time for any small value of $a > 0$ ($\theta \rightarrow 0$). In Fig.\,\ref{1D_effctacce}, $\cos(\theta_0 e^{-at})$ has been plotted as function of $t$ for different value of $a$.  For higher value of $a$ we can see $\cos(\theta_0 e^{-at})$ reaching unity very early in time. After reaching unit value, the walker will continue to spread with maximum velocity. This also gives us a range of values $a$ to model and study the effect of different rate acceleration on the quantum walk dynamics. 

Fig.\,\ref{1D_prob} represents the probability distribution for single-particle accelerated quantum walk in one dimension for different value of $a$ and $\theta_0$. As the value of $a$ increases, the spread in the position space increases. This can be understood by noting a faster decrease in value of $\theta$ with increase in the value of  $a$ and $t$,  that is, $\theta(a, t) \longrightarrow 0$ spreads probability distribution over larger space. The value $\theta_0$ sets the minimum spread (non-accelerated) a walker can achieve.  In Figs.\,\ref{1D_sdsteps} and \ref{1D_sdacce} standard deviation ($\sigma$) as function of number of steps ($t$) and the acceleration ($a$) for different values of $\theta_0$ is shown.  A significant increase in $\sigma$ with increase in $a$ is seen until it attains a maximum spread bounded by $\sigma = t$ and $\theta_0$ sets the minimum $\sigma$.
\begin{figure}[h!]
\centering
\includegraphics[width = \linewidth]{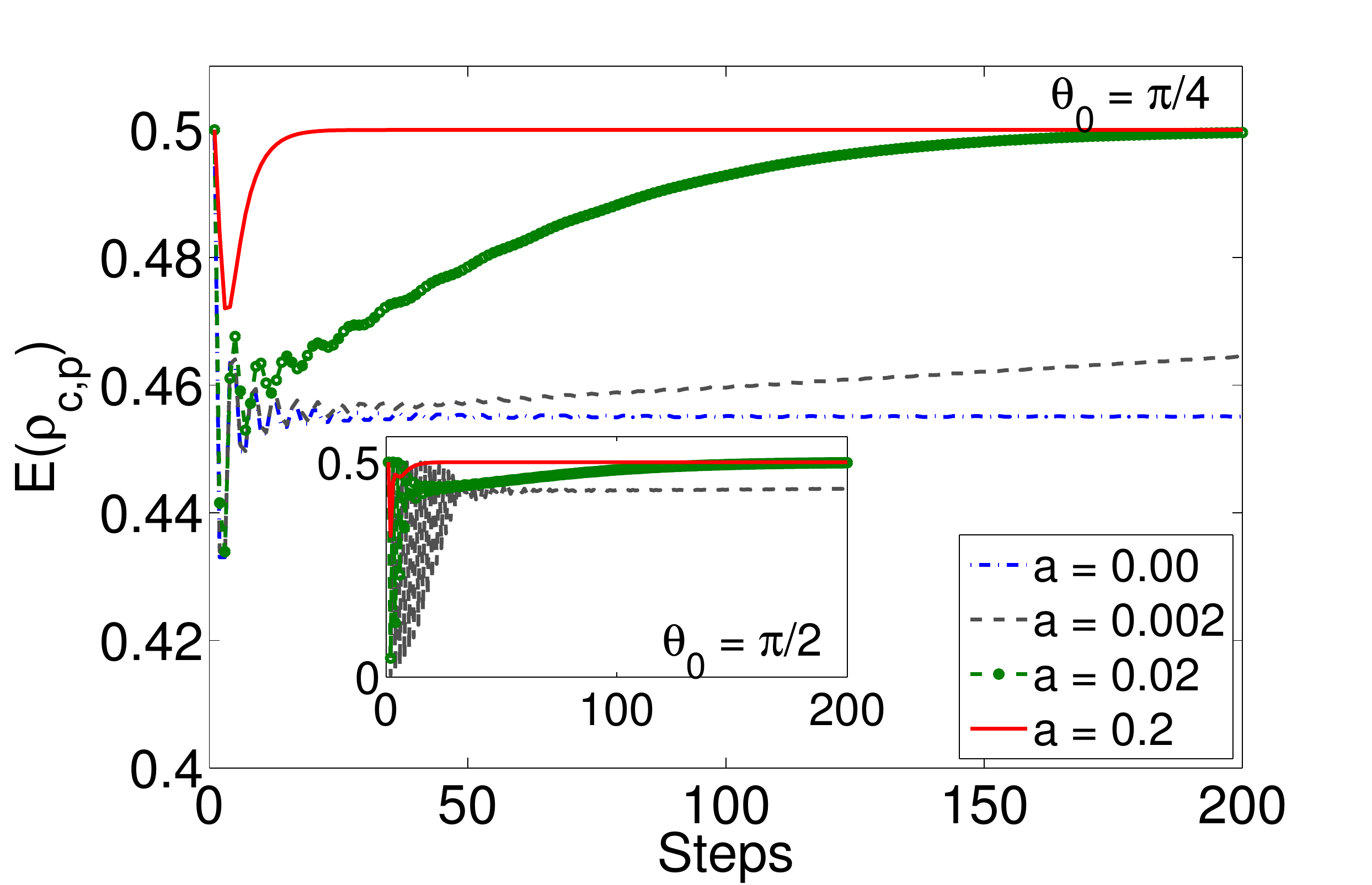}
\caption{Negativity of an one-dimensional accelerated quantum walk with $\theta(a, t) = \theta_0 e^{-at}$ for different value of $a$ and $\theta_0$ as function of steps ($t$). $a=0$ corresponds to evolution with constant velocity, an homogeneous evolution and the negativity is minimum for this case and reaches a study value below the maximum. The initial state is $\ket{\Psi_{in}} = \frac{1}{\sqrt 2}(\ket{\uparrow} + \ket{\downarrow}) \otimes \ket{x = 0}$. Negativity is  shown for two values of $\theta_0 = \pi/4$ and $\pi/2$ (inset). With increase in $a$ an increase in steps and an increase in negativity is seen. In the inset entanglement is plotted only for $a \neq 0$  values. For $a=0$ we will only see the oscillation between the maximum and minimum values.}
\label{1D_aqwNeg}
\end{figure}

\noindent
{\it Entanglement --} In quantum walk dynamics the superposition and quantum interference results in quantum correlation between the particle and the position space~\cite{RSCB, CGB, KRS,SC}. Quantum correlations in quantum walks using different forms of measure has been investigated~\cite{ BBB, ASRD,CLXG} and some interesting results like enhancement of entanglement with temporal disordered coin operations have been reported ~\cite{CMC, VAR13}. In this section we will look into the effect of acceleration on entanglement generated in the system, between particle and position space. This will help us in understanding the entanglement generation between two-particles and its survival during quantum walk of two accelerated particle in the following sections. We will use negativity as a quantifying measure of entanglement which is given by,
\begin{equation}
E(\rho_{c,p}) \equiv \mathcal{N}_{c, p} = \sum_i \frac{(|\lambda_i| - \lambda_i)}{2}
\label{neg1}
\end{equation}
where, $\lambda_i$ is the eigenvalue of the partial transpose of the density matrix $\rho$.  Negativity is chosen as a measure to quantify entanglement because of its validity even when we have a mixed state, that is, when we trace out position space and look at entanglement between the two particles in two-particle  system we study in the following sections. Fig. \ref{1D_aqwNeg} shows the negativity between the particle and position space for different value of acceleration parameter $a$ in the coin operation parameter $\theta$. It is compared with the negativity generated in standard discrete-time quantum walk  (homogeneous) which can be recovered by substituting $a=0$ and $\theta_0 = \pi/4$.  The maximum value of the negativity for the formula we have used in Eq.\,(\ref{neg1}) will be 0.5. For an accelerated quantum walk we can see a study increase in negativity with increase in number of steps. For higher value of $a$, negativity increases with number of steps and quickly reaches a maximum value and remains maximum with further increase in steps. For smaller value of $a$, a very slow increase in negativity as function of number of steps is seen. In general, one can see a significant increase in negativity with increase in acceleration and with increase in number of steps (time) until it reaches a maximum value.

Increase in standard deviation with acceleration happens due to the transition of quantum coin operation towards identity operator with time and for higher value $a$ coin evolution operator tends towards identity faster. This implies that with $a>0$ a particle which spreads in position space localizes the two states $|\uparrow\rangle$ and $|\downarrow \rangle$ to bimodal form and move away from each other. 

The increase in negativity seen with time dependent coin operation causing acceleration allies well with the earlier results of increase in entanglement between particle and position space due to random coin operation for each time~\cite{CMC, VAR13}. Thus, one can say that the enhancement of quantum correlation (negativity or entanglement entropy) is in general due to the time dependent quantum coin operation and not due to a specific effect of randomness in time. 

\subsection{\label{subsec2} Two-particle} 

Understanding and modelling the dynamics of interacting quantum particles has been one of the prime interest of study for over decades now\,\cite{LSM, CHN}. In particular, with advances in quantum information theory, entanglement between the interacting particles in many body physics has received special interest\,\cite{CB,AFOV,BW,KD}. Here we present a simple model to study the entanglement between the two interacting particle and the effect of acceleration using accelerated two-particle discrete-time quantum walk. 

\begin{figure}[h!]
\centering
\includegraphics[width = \linewidth]{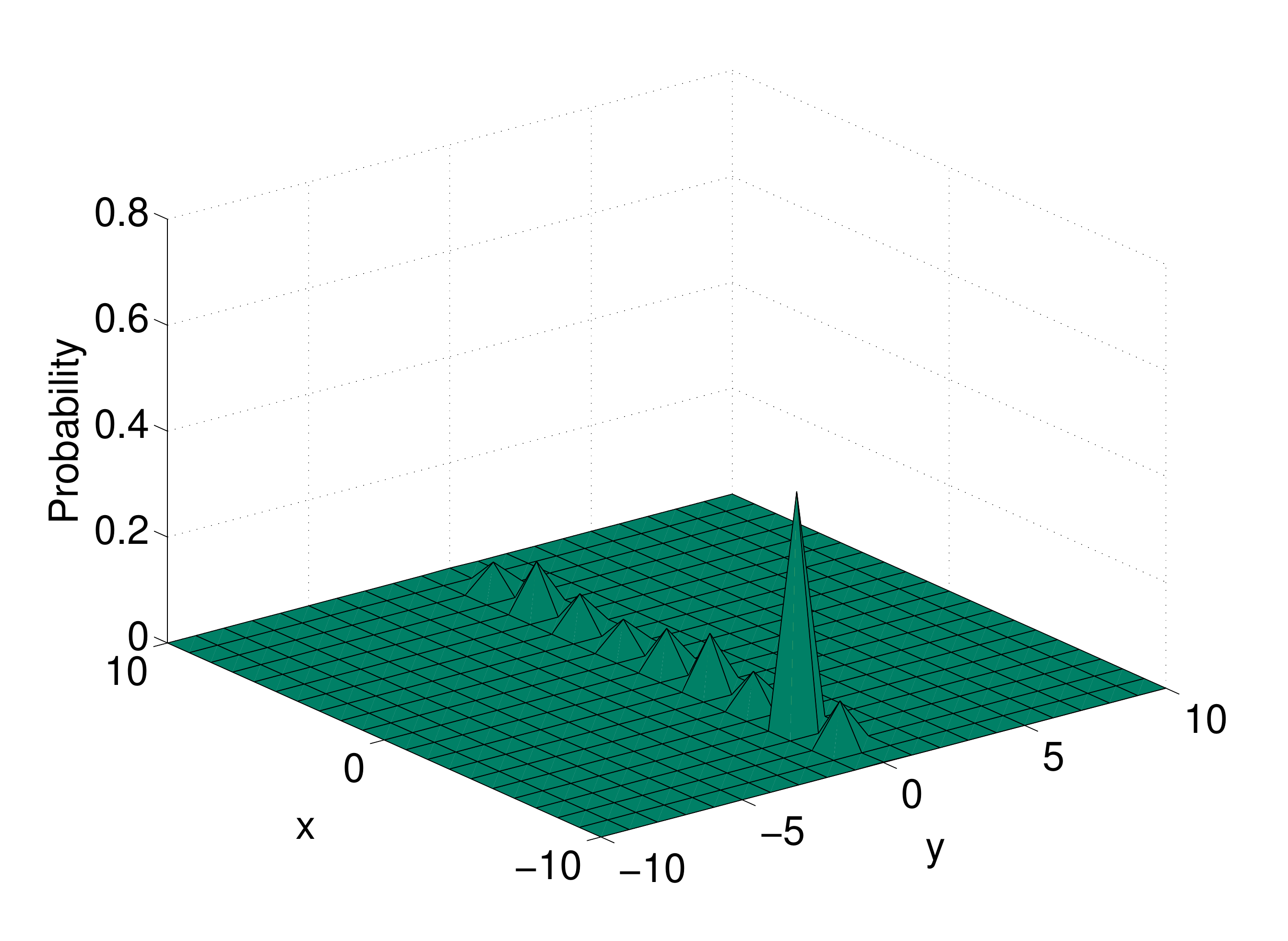}
\caption{ Probability distribution for two-particle discrete-time quantum walk in two-dimensional position space when the initial state is $\ket{\Psi_{in}} = \ket{\uparrow\uparrow} \otimes \ket{x = 0} \otimes \ket{y=0}$ for $\theta_0 = \pi/4$.}
\label{OneState}
\end{figure}

\begin{figure}[h!]
\centering
\includegraphics[width = \linewidth]{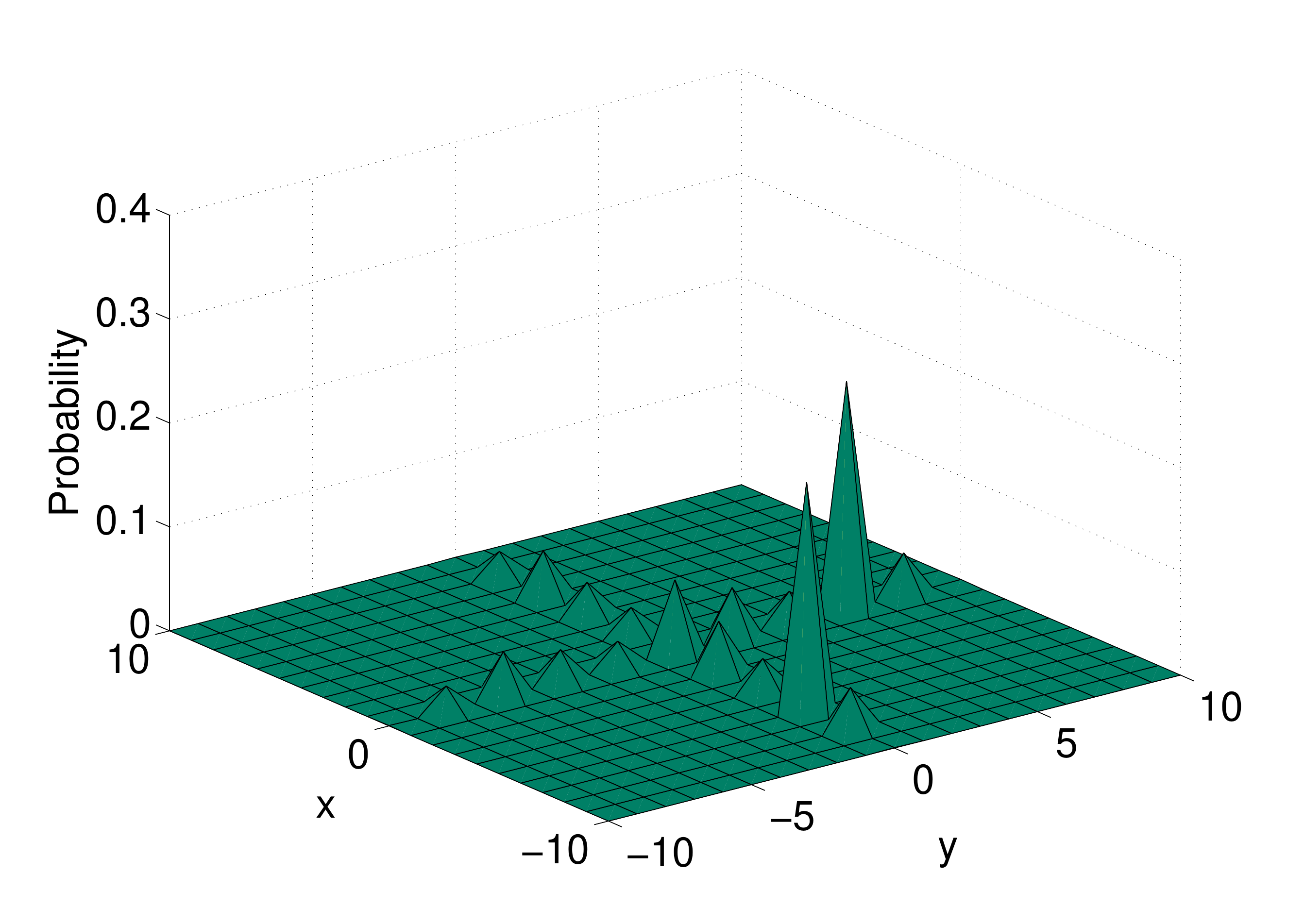}
\caption{ Probability distribution for two-particle discrete-time quantum walk in two-dimensional position space when the initial state is $\ket{\Psi_{in}} = \frac{1}{\sqrt{2}} (\ket{\uparrow\uparrow} + \ket{\uparrow\downarrow}) \otimes \ket{x = 0} \otimes \ket{y=0}$ for $\theta_0 = \pi/4$.}
\label{TwoState}
\end{figure}

\begin{figure}[h!]
\centering
\includegraphics[width = \linewidth]{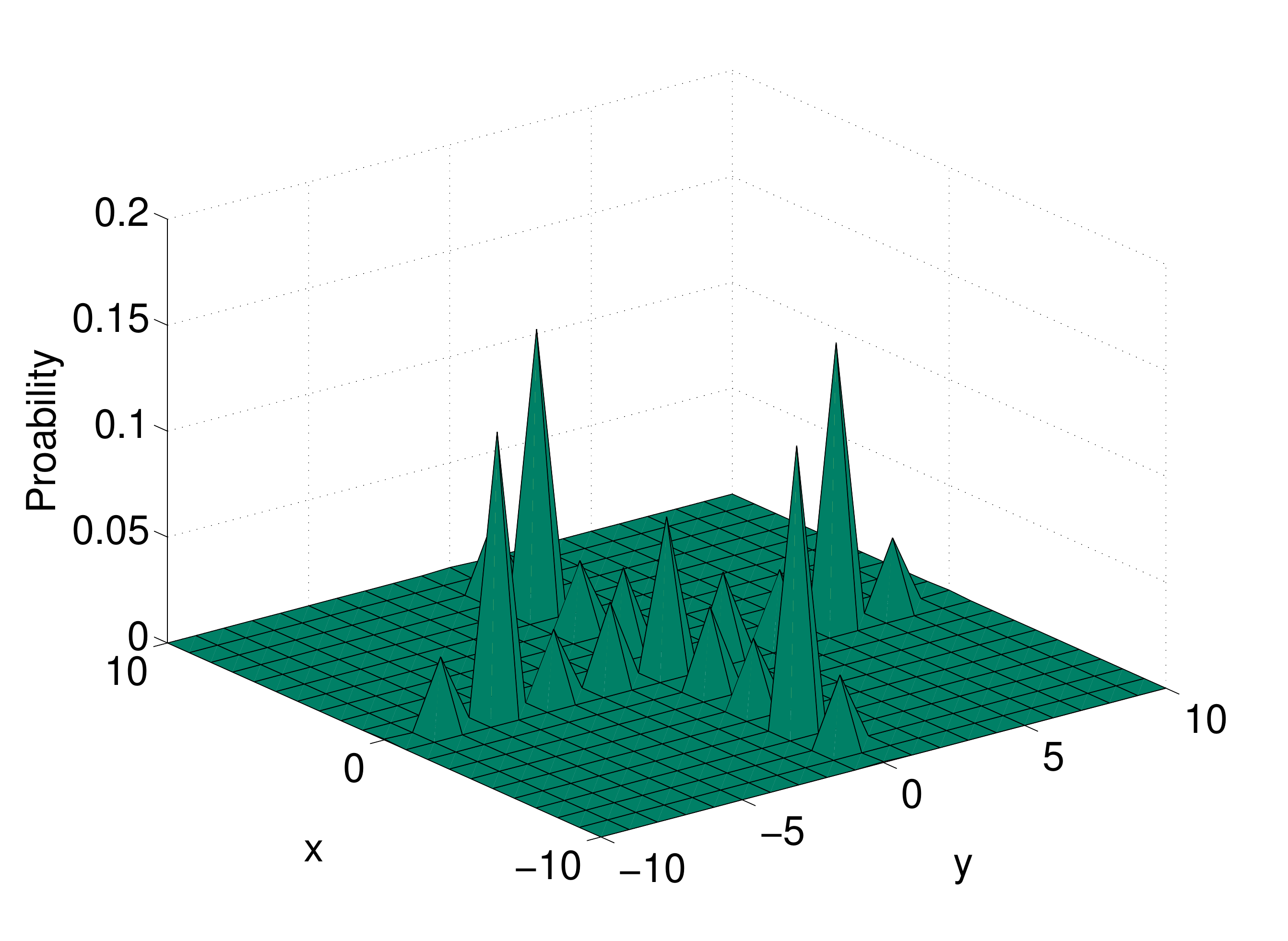}
\caption{ Probability distribution for two-particle discrete-time quantum walk in two-dimensional position space when the initial state is $\ket{\Psi_{in}} = \frac{1}{2} (\ket{\uparrow\uparrow} + \ket{\uparrow\downarrow} + \ket{\downarrow\uparrow} + \ket{\downarrow\downarrow}) \otimes \ket{x = 0} \otimes \ket{y=0}$ for $\theta_0 = \pi/4$.}
\label{State}
\end{figure}

The two-particle discrete-time quantum walk is defined on a Hilbert space $\mathcal{H} = \mathcal{H}_{c_1} \otimes \mathcal{H}_{c_2} \otimes \mathcal{H}_{p_{x}} \otimes \mathcal{H}_{p_{y}}$, where $\mathcal{H}_{c_1}$ and $\mathcal{H}_{c_2}$ are the Hilbert space  composed of the two internal degrees of freedom of the two-particles, respectively. The particle Hilbert space $\mathcal{H}_{c_j} = span\left\lbrace \ket{\uparrow}, \ket{\downarrow} \right\rbrace$  where $j= 1,2$ and position Hilbert space $\mathcal{H}_{p_x} = span\left\lbrace \ket{x} \right\rbrace , x \in \mathbb{Z}$ and $\mathcal{H}_{p_y} = span\left\lbrace \ket{y} \right\rbrace , y \in \mathbb{Z}$ represents the position basis states in two dimension. The evolution operator for a two-particle quantum walk in two-dimensional space is defined using an unitary interacting operator $C_{\theta}$ followed by position shift operator $S_{(x,y)}$. The interacting operator on particle Hilbert space is given as,
\begin{align}
\label{intcoin}
C_{\theta_0} &= \Bigg (\cos(\theta_0) (\mathbb{I}_{2} \otimes \mathbb{I}_2) - i\sin(\theta_0) (\sigma_x \otimes \sigma_x) \Bigg )  \nonumber \\      
&= \begin{pmatrix}
\cos(\theta_0) & 0 & 0 & -i\sin(\theta_0) \\
0 & \cos(\theta_0) & -i\sin(\theta_0) & 0 \\
0 & -i\sin(\theta_0) & \cos(\theta_0) & 0 \\
-i\sin(\theta_0) & 0 & 0 & \cos(\theta_0)
\end{pmatrix}.   
\end{align}
Here $\sigma_x$ is the $x-$ Pauli matrix. The $C_{\theta_0}$, decides the amplitudes of the  occurrence of each allowed states  and thus acts as two-particle quantum coin operation. It invokes Ising interaction with Hamiltonian $H = \sigma_x \otimes \sigma_x$ between the two particles.  The position shift operator $S_{(x,y)}$, takes the form, 
 \begin{align}
S_{(x,y)} \equiv& \sum_{(x,y)} \left[ \ket{\uparrow \uparrow}\bra{\uparrow \uparrow} \otimes \ket{\textit{x}-1}\bra{\textit{x} } \otimes \ket{\textit{y}}\bra{\textit{y} } \right.\nonumber \\
& + \ket{\uparrow \downarrow}\bra{\uparrow \downarrow} \otimes \ket{\textit{x}}\bra{\textit{x} } \otimes \otimes \ket{\textit{y}+1}\bra{\textit{y} } \nonumber \\
& + \ket{\downarrow \uparrow}\bra{\downarrow \uparrow} \otimes \ket{\textit{x}}\bra{\textit{x}} \otimes \ket{\textit{y}-1}\bra{\textit{y} } \nonumber \\
& \left. +\ket{\downarrow \downarrow}\bra{\downarrow \downarrow} \otimes \ket{\textit{x}+1}\bra{\textit{x}}  \otimes \ket{\textit{y}}\bra{\textit{y} } \right]
\end{align}
which evolves the particles into superposition of position space. The unitary evolution operation for each step of two-particle walk is given by,
\begin{equation}
\label{2peo}
W(\theta_0) \equiv S_{(x,y)} \Big (C_{\theta_0} \otimes \mathbb{I}_N \Big )
\end{equation}
where $\mathbb{I}_N$ is the identity operator on the position space of length $N$.  After $t$-time steps, state of the system 
\begin{align}
\ket{\Psi_t^{2p}} &= W(\theta_0)^t \ket{\Psi_{in}^{2p}} \nonumber \\
 &= \sum_{(x,y)} \Big ( \mathcal A_{x, t}|\uparrow \uparrow \rangle +  \mathcal B_{x, t}|\downarrow \downarrow \rangle \Big) \otimes |x,y_0 \rangle \nonumber \\
 &+ \Big(\mathcal C_{y, t}|\uparrow \downarrow \rangle + \mathcal D_{y, t}|\downarrow \uparrow \rangle \Big ) \otimes |x_0,y \rangle.
\end{align}
\begin{figure}[h!]
\centering
\includegraphics[width = \linewidth]{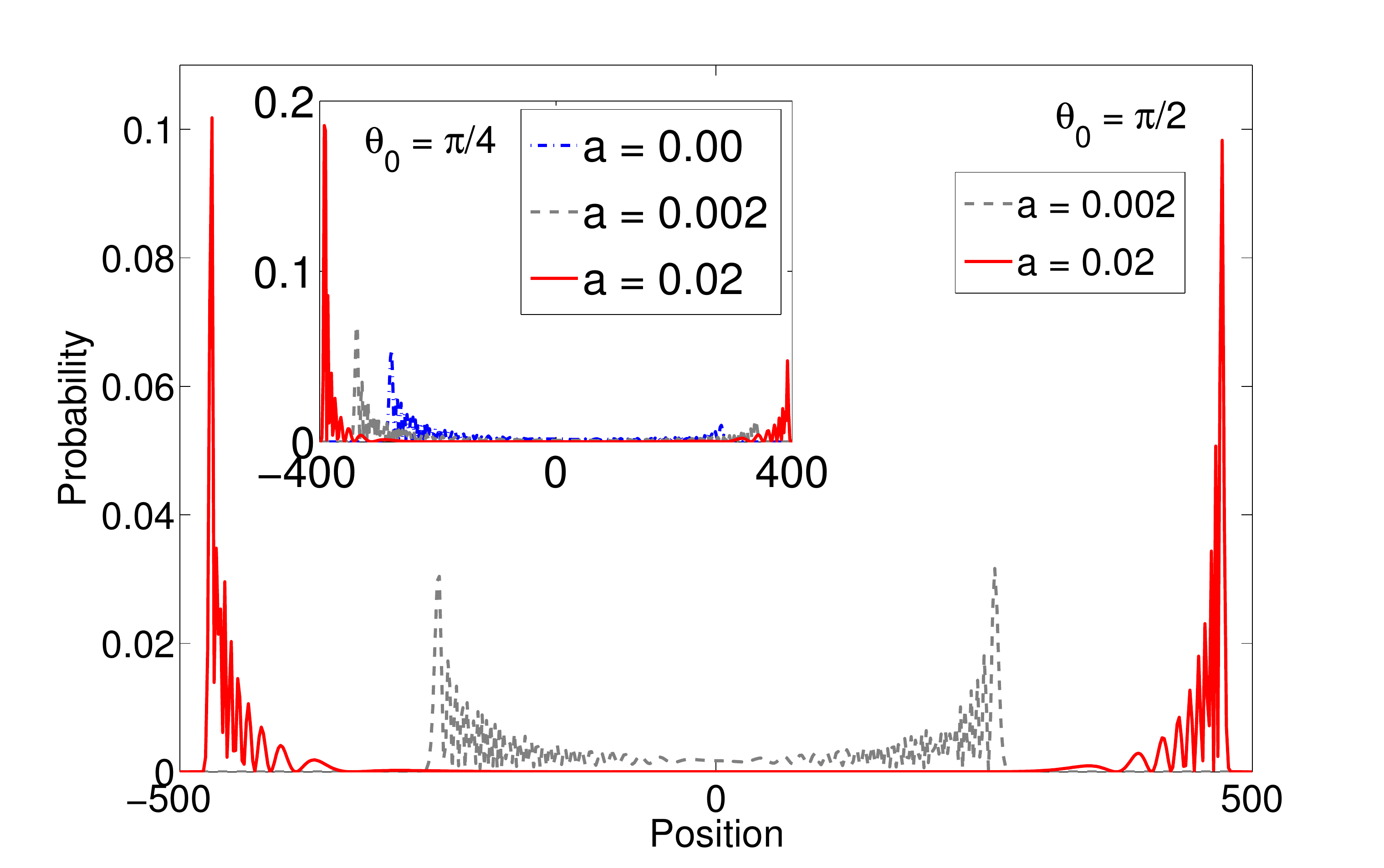}
\caption{Probability distribution of two-particle accelerated quantum walk in one-dimension with $\theta(a, t) = \theta_0 e^{-at}$ for different value of $a$ after $t $ steps of walk. For $\theta_0 = \pi/2$,$t = 500$ and for $a=0$ which corresponds to an homogeneous evolution, does not give any spread in the position space therefore is not shown here but with increase in $a$, spread in position space increases symmetrically for unentangled initial state. In the inset we show probability distribution for $\theta_0 = \pi/4$ for $t=400$ steps for different values of $a$ and here we see asymmetric probability distribution for unentangled initial state. The initial state is $\ket{\Psi_{in}} = \ket{\uparrow\uparrow} \otimes \ket{x = 0} \otimes \ket{y=0}$.}
\label{Prob_AQW}
\end{figure}
For three different initial state in the form, $|\Psi_0^{2p}\rangle = |\psi_{c} \rangle \otimes |x_0,y_0 \rangle$, the probability distribution after 10 steps in two-dimensional position space is shown in Fig. \ref{OneState}, \ref{TwoState} and \ref{State}. We can see that the probability distribution in x-direction and y-direction are independent of each other and are dependent on initial coin state. If the initial state of the two particle is in one of the basis states  $| \uparrow \uparrow \rangle \otimes |x=0\rangle \otimes |y=0\rangle $ or $ | \uparrow \downarrow \rangle \otimes |x=0\rangle \otimes |y=0\rangle $ the state after $t$ step will evolve only in configuration of two of the basis states and one-dimensional space,
\begin{align}
\ket{\Psi_t^{2p}}& = W(\theta_0)^t  | \uparrow \uparrow \rangle \otimes |x=0\rangle \otimes |y=0\rangle   \nonumber \\
&= \sum_{x}  \Big ( \mathcal A_{x, t}|\uparrow \uparrow \rangle +  \mathcal B_{x, t}|\downarrow \downarrow \rangle  \Big ) \otimes |x \rangle \otimes |y = 0 \rangle,\\
 \ket{\Psi_t^{2p}}  & = W(\theta_0)^t  | \uparrow \downarrow \rangle \otimes |x=0\rangle  \otimes |y=0\rangle \nonumber \\
&= \sum_{y}  \Big ( \mathcal C_{y, t}|\uparrow \downarrow \rangle + \mathcal D_{y, t}|\downarrow \uparrow \rangle  \Big ) \otimes |x = 0 \rangle \otimes |y \rangle.
 \end{align}
This initial state and the choice of coin and shift operator we have chosen ensures the fermionic and bosonic nature of the wave function as the symmetry and anti-symmetry is introduced in the spin state. The initial position space is also symmetric i.e., $|x=0\rangle \otimes |y=0\rangle$. Our interest here is to begin with an unentangled two-particle state and explore the dynamics that can entangle and further investigate its survival with time. Therefore, for all the two-particle dynamics we will hereafter set the initial state to be, $\ket{\Psi_{in}^{2p}} = | \uparrow \uparrow \rangle \otimes |x=0\rangle \otimes |y=0\rangle$ and this should hold good even for the initial two particle state $| \uparrow \downarrow \rangle$. For this configuration we can note that the probability amplitude spreading of two-particles will evolve together just like a single-particle quantum walk.  However, this simple evolution will still allow us to explore the interesting dynamics resulting in entanglement generation between these two particles evolving together.
 
Acceleration in the two-particle quantum walk is introduced by replacing $\theta_0$ in the interaction operator, Eq.\,(\ref{intcoin}) with $\theta(a, t) = \theta_0 e^{-at}$ where $a$ is the parameters which set the acceleration rate. Fig.\,\ref{Prob_AQW} shows the probability distribution of the two-particle accelerated quantum walk with different values of acceleration parameter $a$ and as the acceleration in the system increases the spread of the probability amplitude in position space also increases.

 \begin{figure}[h!]
\centering
\includegraphics[width = \linewidth]{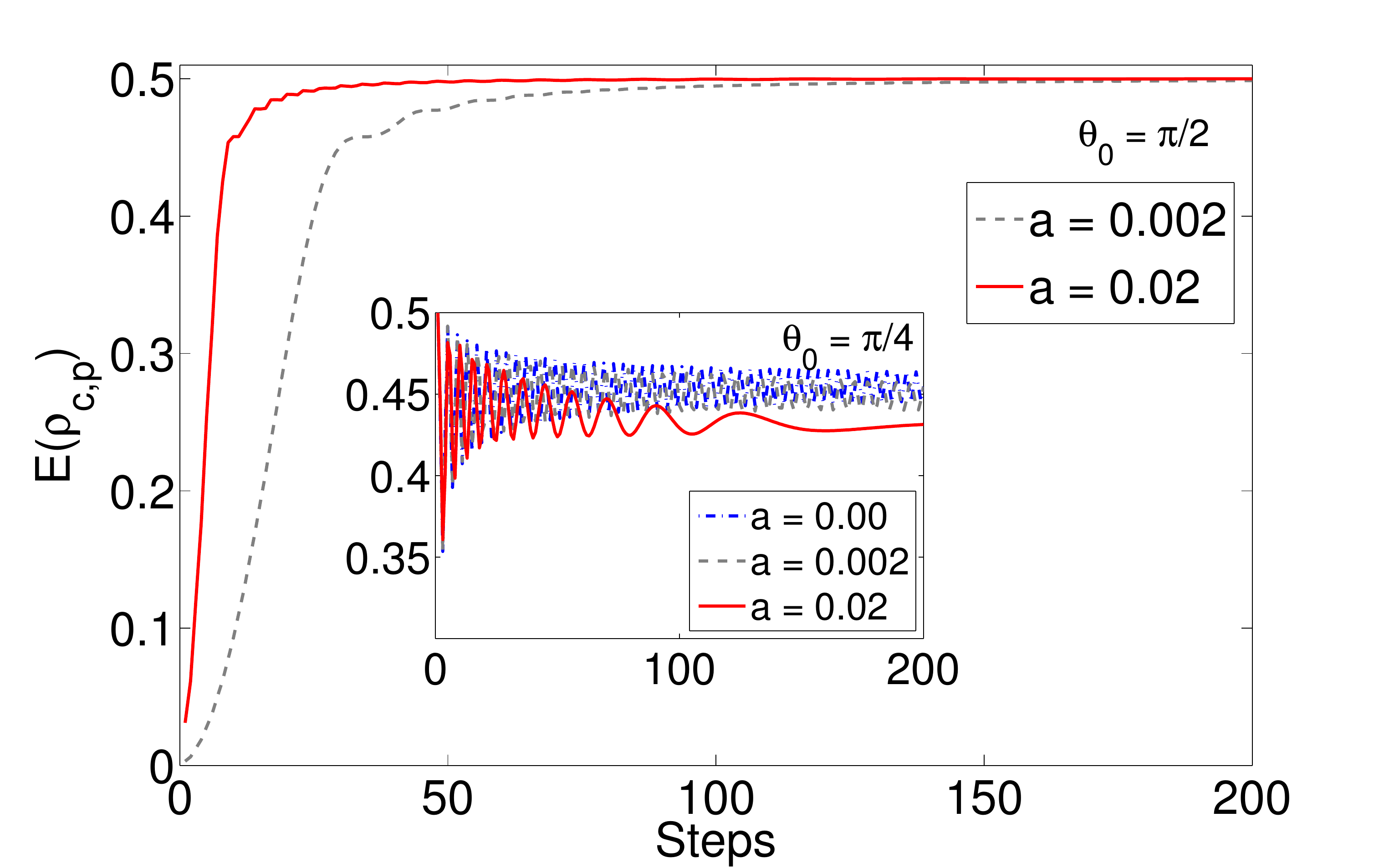}
\caption{Negativity of two-particle accelerated quantum walk in one-dimension position Hilbert space $\mathcal{H}_{p_x}$ between position and particle space with $\theta(a, t) = \theta_0 e^{-at}$ for different value of $a$ and $\theta_0$ as function of steps ($t$). Probability distribution in Hilbert space $\mathcal{H}_{p_y}$ is zero for the given initial state. $a=0$ corresponds to evolution with constant velocity, an homogeneous evolution. The initial state is  $\ket{\Psi_{in}} = \ket{\uparrow\uparrow} \otimes \ket{x = 0} \otimes \ket{y=0}$. Negativity is  shown for two values of $\theta_0 = \pi/2$ and $\pi/4$ (inset). With increase in $a$ an increase in negativity is seen with time for $\theta_0 = \pi/2$ but a decrease in negativity is seen with time for $\theta_0 = \pi/4$.}
\label{Entangle_QW}
\end{figure}
\begin{figure}[h!]
\centering
\includegraphics[width = \linewidth]{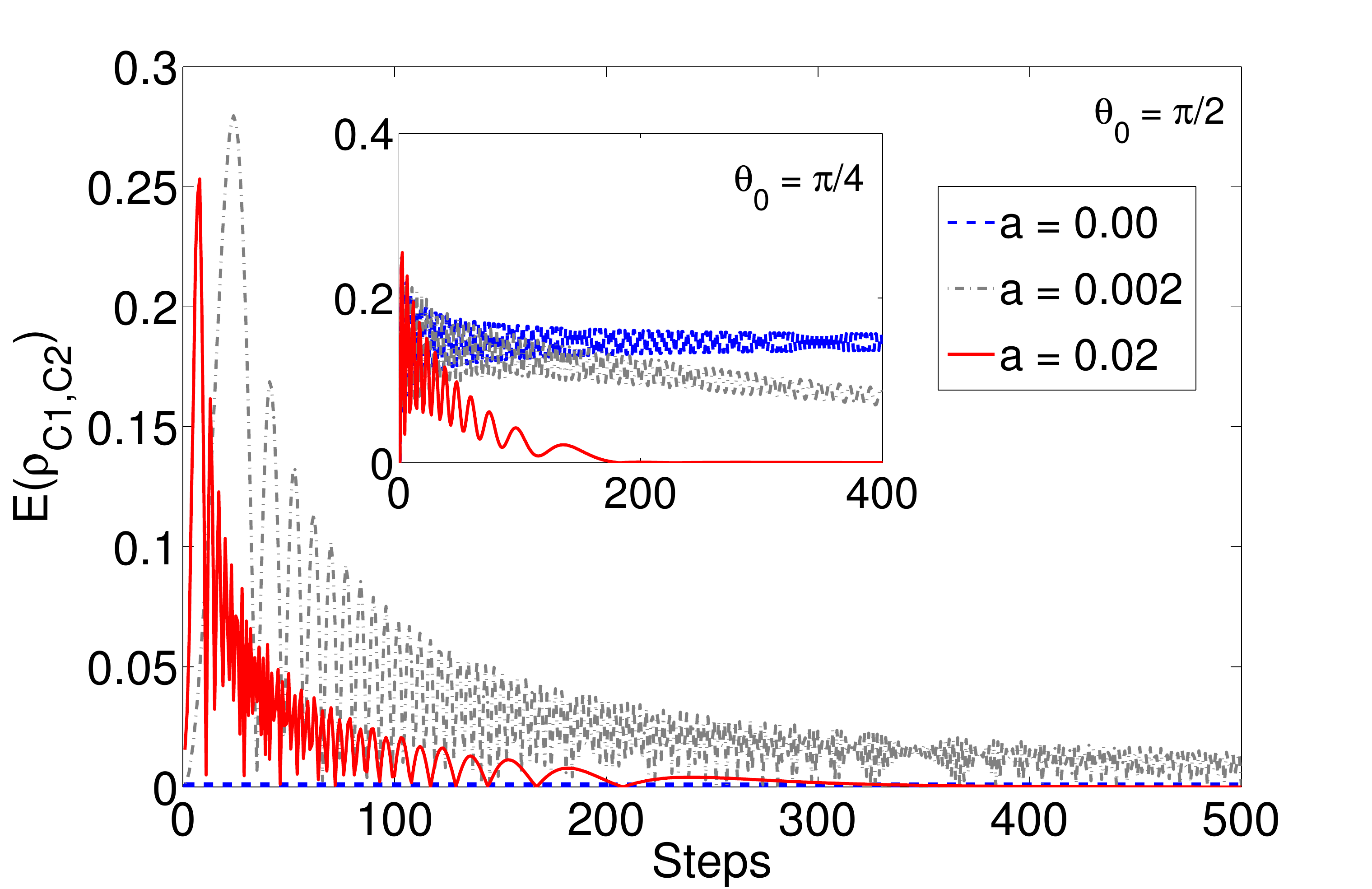}
\caption{Negativity of two-particle accelerated quantum walk between particle space with $\theta(a, t) = \theta_0 e^{-at}$ for different value of $a$ and $\theta_0$ as function of steps ($t$). $a=0$ corresponds to evolution with constant velocity, an homogeneous evolution. The initial state is  $\ket{\Psi_{in}} = \ket{\uparrow\uparrow} \otimes \ket{x = 0} \otimes \ket{y=0}$. Negativity is  shown for two values of $\theta_0 = \pi/2$ and $\pi/4$ (inset). With increase in $a$ a decrease in negativity is seen with time.
}
\label{Nega_QW}
\end{figure}

{\it Entanglement -} For a one-particle quantum walk we only had two the Hilbert space associated with the dynamics, particle and position space. Therefore, only entanglement between the particle and position space was calculated. For a two-particle quantum walk we will independently have Hilbert space of two particles and both the dimension of the position space.
Thus, one can calculate the entanglement between the two-particle and position space, and entanglement between just the two particle after tracing out the position space.  

Entanglement between the two-particle walker and position Hilbert space is given by the negativity between the two-particle (coin) and position Hilbert space, 
 \begin{equation}
 E(\rho_{c,p}) \equiv \mathcal{N}_{\rho_{c,p}} =  \frac{1}{2}\sum_{j} (|\zeta_{j}|- \zeta_{j})
 \end{equation}
where, $\zeta_{j}$ with $j \in \mathbb{Z}$ are the eigenvalues of partial transpose of density matrix $\rho$ in Hilbert space $\mathcal{H} = \mathcal{H}_{c1} \otimes \mathcal{H}_{c2} \otimes \mathcal{H}_{p_x}$.

Entanglement between the two-particle Hilbert space is given by the negativity in the particle Hilbert space after tracing out the position Hilbert space from the total density matrix $(\rho = \ket{\psi}\bra{\psi})$. That is, negativity of the mixed state of two-particle Hilbert space after tracing out the position Hilbert space $\rho_c = tr_p(\rho)$ is given by,
  \begin{equation}
 E(\rho_{C1,C2}) \equiv \mathcal{N}_{\rho_{C1,C2}} =  \frac{1}{2}\sum_{i} (|\lambda_{i}|- \lambda_i)
 \end{equation}
 where, $\lambda_i$ with $(i = 1,2,3,4)$ are the eigenvalues of partial transpose of density matrix $\rho_c$. To study the entanglement between the  two particles in accelerated quantum walk we will choose one of the basis of coin state as initial coin state and hence the walk can be reduced to one-dimensional position-space accelerated quantum walk. 

In Fig.\,\ref{Entangle_QW}, negativity between the Hilbert space of the two quantum walker and one of the position Hilbert space $\mathcal{H}_{p_x}$ in which it evolves as function of time ($t$) is shown for different value of $a$ when $\theta_0 = \pi/2$ and $\pi/4$. As the probability distribution in $y$-direction is zero for the initial state $\ket{\Psi_{in}} = \ket{\uparrow\uparrow} \otimes \ket{x = 0} \otimes \ket{y=0}$ we have ignored the $\mathcal{H}_{p_y}$. For an accelerated walker for any value $a$ when $\cos ( \theta(a, t))$ goes from $0$ to $1$ with time $t$ ($\theta_0 = \pi/2$) we see an increase in negativity with time and maximum value of negativity is reached by an walker with higher value of $a$.  However, for an accelerated walker  with an offset $\theta_0 = \pi/4$  ($\theta_0 < \pi/2$) the value of $\cos ( \theta(a, t))$ increase with a small gradient and the effect of acceleration does not last long in the dynamics. Therefore, we don't see any noticeable increase in entanglement when compared to quantum walk without acceleration.   Heuristically we can say that a gradient in acceleration before it is saturated to maximum value plays an important role in generation of maximum entanglement between the two-particle and the position space. 

In Fig.\,\ref{Nega_QW}, negativity just between the two particles after tracing out the position space as function of time ($t$) for different value of $a$ when $\theta_0 = \pi/2$ and $\pi/4$ is shown. For non-accelerated quantum walk ($a=0$) when $\theta_0 = \pi/2$ we do not see any entanglement being generated between the two particles. However, for small value of $a$ we can notice generation of entanglement between the two particles and after a long time evolution it fades off. With increase in $a$, entanglement generated decreases and fades off faster with time. When the offset of $\theta(a, t)$ is set with $\theta_0 = \pi/4$ and $a=0$ the coin operator Eq.\,\ref{intcoin} itself will acts as an entangling operator and it entangles the two particles instantaneously and remains entangled. However, with increase in $a$ the instantaneous entanglement generated fades off faster dis-entangling the two particles. 

\begin{figure}[h!]
\centering
\includegraphics[width = \linewidth]{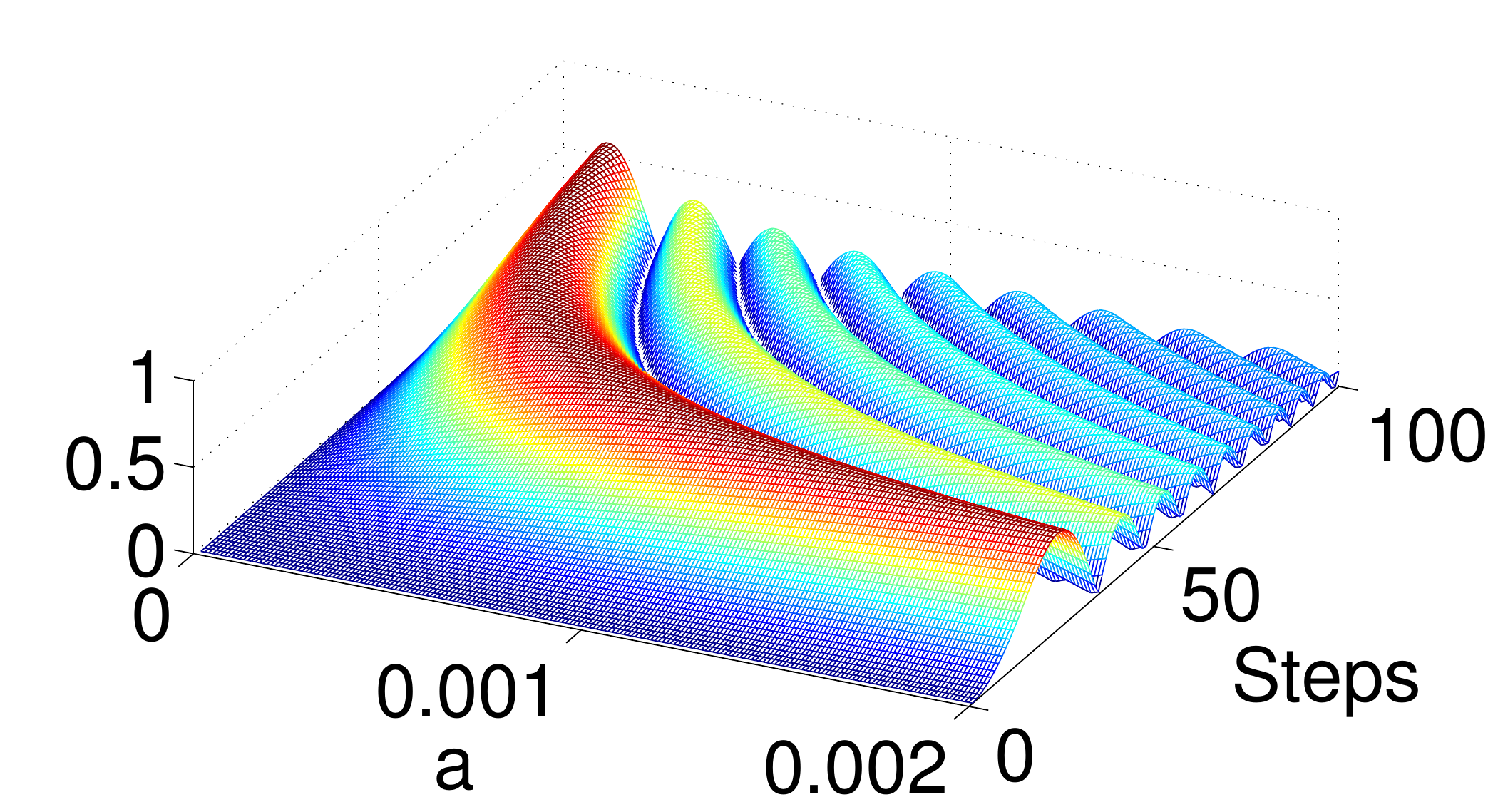}
\caption{Negativity between the two particles after tracing out the position space using $\theta(a, t) = \theta_0 e^{-at}$ as a function of both $a$ and $t$ when $\theta_0 = \pi/2$. $a=0$ corresponds to evolution with constant velocity, an homogeneous evolution. The initial state is  $\ket{\Psi_{in}} = \ket{\uparrow\uparrow} \otimes \ket{x = 0} \otimes \ket{y=0}$. For small value of $a$ one can see the entanglement generated with accelerated quantum walk does not fade off with time (steps). 
}
\label{Nega_2QW}
\end{figure}

To have a further clarity on the generation of entanglement between the two particles when accelerated, in Fig.\,\ref{Nega_2QW} we have plotted negativity as function of both, $a$ and $t$ when $\theta_0=\pi/2$. We can arrive at the following conclusion : for extremely small value of $a$ the entanglement between the two particles picks up slowly and will remain entangled with time. With increase in both $a$ and $t$ simultaneously, entanglement fades off faster with time. 

Our choice of initial state and interacting coin operation in this subsection on two-particle quantum walks reproduces the dynamics identical to the single-particle quantum walk. Therefore, it is appropriate to draw connection of two-particle quantum walk described in this section with the established connection of single-particle quantum walk with the Dirac equation\,\cite{DAM, bia94, Str06, CBS10, Cha13, AP14, Per16, MC16}.  Parameter $\theta$ in the quantum walk coin operation or in the operations that describe Dirac cellular automata has been effectively mapped to the mass and velocity of Dirac particle, that is, $m \propto \sin(\theta)$ and $c \propto \cos(\theta)$.  This allows us to connect the role of mass in generation of entanglement between the two particles of same mass under quantum walk evolution. 
When $\theta$ is a function of $a$ and $t$ we see that small value of $a$ corresponds to larger mass of the particle. Thus, higher the mass, slower is the spread and higher is the entanglement generation.

\section{Disorder and localization} \label{sec3}

\subsection{Single-particle}


Disorder in discrete-time quantum walk has been studied by replacing $\theta_0$ in the quantum coin operation with a random value $ 0 \leq \theta_j  \leq \pi$ in position ($j=x$) or time ($j = t$)\,\cite{CMC,CM, SC}.  In an alternative approach, disorder has been introduced in the form of random phase operation along with the coin operation with fixed $\theta_0$\,\cite{VFQ,SD}. In order to introduce disorder in discrete-time quantum walk by retaining the parameter that will accelerate the walker, we will use the later approach and introduce disorder in phase operator $\Phi(\phi)$ given by,
\begin{equation}
\Phi_{\phi} = \begin{pmatrix}
1 & 0\\
0 & e^{i\phi}
\end{pmatrix} \otimes \sum_{x} \ket{x}\bra{x}.
\end{equation} 
The effective coin operator without disorder and acceleration along with the phase operator will be $B(\theta_0, \phi) = \Phi(\phi)C(\theta)$ which in matrix form will be given by,
\begin{equation} \label{eqEvol}
B(\theta_0, \phi) = \begin{pmatrix}
 \cos(\theta_0) & -i \sin(\theta_0) \\
-i e^{i \phi} \sin(\theta_0) & e^{i \phi} \cos(\theta_0) 
\end{pmatrix} \otimes \sum_{x} \ket{x}\bra{x}
\end{equation}
and the evolution operator for each step of the quantum walk will have the form, 
\begin{equation} \label{EV_OP}
W = S_x\Phi(\phi)C(\theta_0) \equiv S_xB(\theta_0, \phi).
\end{equation}
The dynamical expression for the evolution operator with phase is given by,
\begin{align} \label{EqGDSP}
\begin{pmatrix}
\mathcal{A}_{x,t} \\
\mathcal{B}_{x,t}
\end{pmatrix} &= \begin{pmatrix}
\cos(\theta_0) & -i\sin(\theta_0) \\
0 & 0
\end{pmatrix} \begin{pmatrix}
\mathcal{A}_{x-1,t-1} \\
\mathcal{B}_{x-1,t-1}
\end{pmatrix} \nonumber \\
&+ \begin{pmatrix}
0 & 0 \\
-ie^{i\phi}\sin(\theta_0) & e^{i\phi}\cos(\theta_0)
\end{pmatrix} \begin{pmatrix}
\mathcal{A}_{x+1,t-1} \\
\mathcal{B}_{x+1,t-1}
\end{pmatrix}
\end{align}
as $(\psi_{x,t}^{\uparrow}; \psi_{x,t}^{\downarrow}) = (\mathcal{A}_{x,t} ;\mathcal{B}_{x,t})$ are the probability amplitudes. For this given evolution operator, in absence of disorder, the Fourier mode wave like form for probability amplitude $\mathcal {A}_{x, t}$ and $\mathcal{B}_{x, t}$ can in general be written as  $\psi_{x,t} = e^{(-i\omega_1 t + i\kappa_1 x)} $ where $\omega_1$ is the wave frequency and $\kappa_1$ is the wave number. Substituting the Fourier form of the probability amplitude in Eq.\,(\ref{EqGDSP}) and solving for a relation between $\kappa_1$ and $\omega_1$ we get the dispersion relation of the form,
\begin{equation}
\cos(\omega_1 + \phi/2) = \cos(\theta_0)\cos(\kappa_1+ \phi/2).
\end{equation} 
This will give the group velocity of the form,
\begin{align} \label{EqGGV}
v_g = \frac{\cos(\theta_0) \sin(\kappa_1 + \phi/2)}{\sqrt{1 - \cos^2(\theta_0) \cos^2(\kappa_1 + \phi/2)}}
\end{align}
and it implies that for evolution operator with phase, the group velocity depends upon both, the evolution parameters $\theta_0$ and $\phi$. Transfer matrix approach will give an insight into the dependence of the amplitudes on the coin parameter when particle moves from one position to an other position at any given time. The transfer matrix for single-particle discrete-time quantum walk in one dimension for coin operator $B(\theta_0, \phi)$ with the two component field $\Psi_x = (\psi_{\uparrow,x}; \psi_{\downarrow,x-1}) =  (\mathcal{A}_{x}; \mathcal{B}_{x-1})$ in generic form at time $t$ is given by,
\begin{equation} \label{1D_trans}
T_{x} = e^{-i \phi_x/2} \begin{pmatrix}
e^{i(\omega_1 + \phi_x/2}\sec(\theta_{x}) & -ie^{-i\phi_{x}/2}\tan(\theta_{x})\\
ie^{i \phi_x/2} \tan(\theta_{x}) & e^{-i(\omega_1+\phi_{x}/2)}\sec(\theta_{x})
\end{pmatrix}.
\end{equation} 
The state at any position using transfer matrix is given by, $\Psi_{x+1} = T_{x} \Psi_{x}$ and this will have both the parameters $\theta_x$ and $\phi_x$. Therefore, the spatial disorder can be introduced in the system by making either one of the coin parameter $\theta_x$ or $\phi_x$ position dependent with a randomized value. Similarly,  the evolution operator Eq. (\ref{EV_OP}) also depends upon both the parameters $\theta$ and $\phi$ of the coin operator, thus, randomizing any one of them with time will give temporal disorder. For accelerated discrete-time quantum walk with disorder we can keep $\theta (a, t) = \theta_t = \theta_0 e^{-a t}$ to accelerate the walk and introduce disorder by randomizing $\phi$ with position or time for spatial and temporal disorder, respectively. 

\begin{figure}[h!]
\centering
\includegraphics[width = \linewidth]{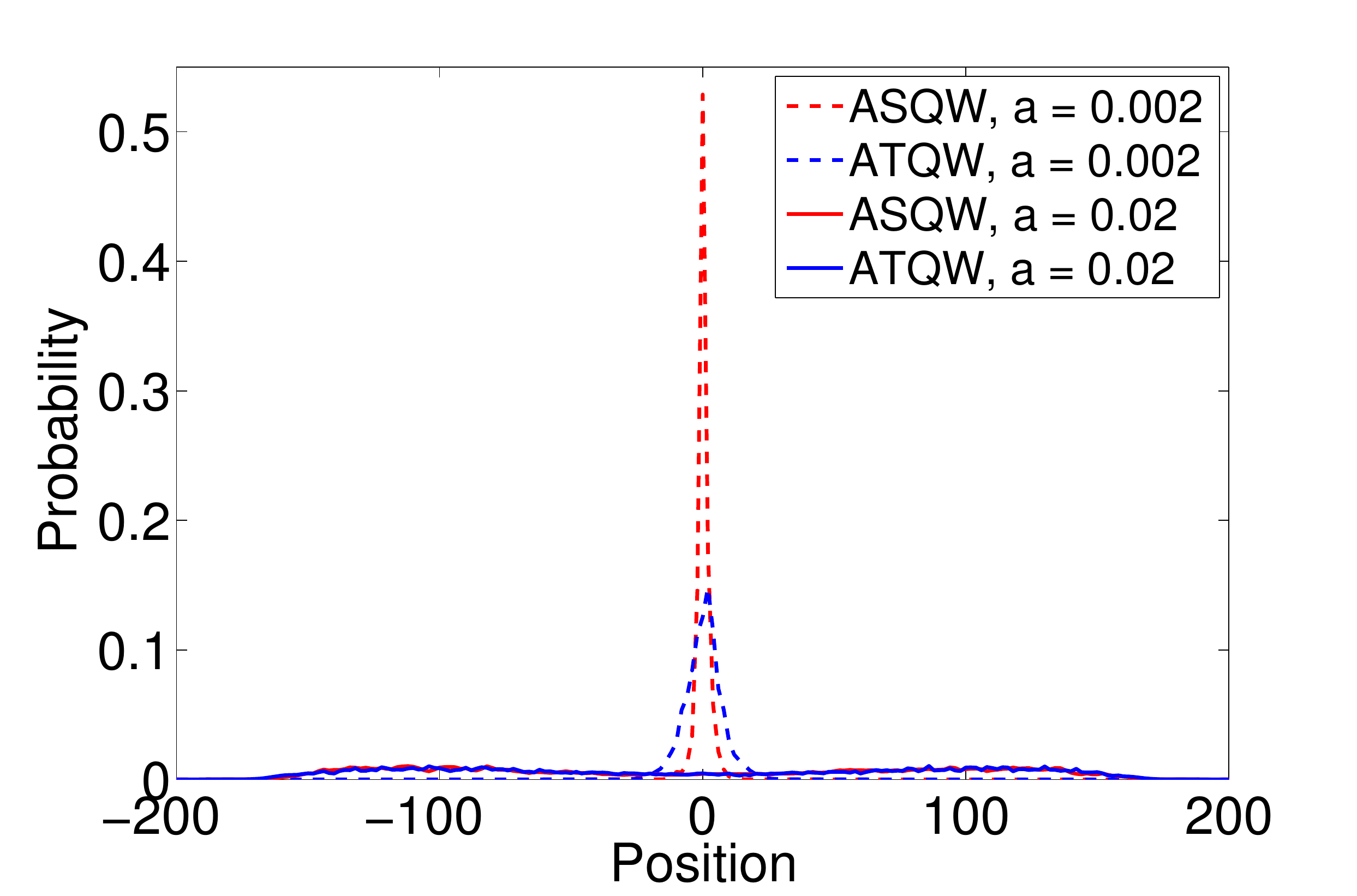}
\caption{Probability distribution for one-dimensional single-particle accelerated quantum walk in disordered system for different value of $a$ averaged over 500-runs for 200-time steps. The initial state is $\ket{\Psi_{in}} = \ket{\uparrow} \otimes \ket{x = 0}$. For higher value of $a$ the distribution is highly delocalized.}
\label{1D_disorder}
\end{figure}

\noindent
{\it Spatial Disorder--}As described above, randomizing $\phi_x$ parameter in coin evolution will lead to spatial disorder and the quantum walk can be accelerated by using coin parameter $\theta_t = \theta_0 e^{-a t}$. Parameter $\phi_x$ is randomly picked for each position from the range $0 \leq \phi_x \leq \pi$. The state after time $t$ with spatial disorder in accelerated discrete-time quantum walk is given by,
       \begin{align}
        \ket{\Psi(t)}_S &= W_x(\theta_{t},\phi'_x)...W_x(\theta_{2},\phi'_x)W_x(\theta_1, \phi'_x) \ket{\Psi_{in}} \nonumber \\
        &= (\prod_{t=1}^{T} W_x(\theta_t, \phi'_x)) \ket{\Psi_{in}} \nonumber \\
        &= (\prod_{t=1}^{T} S_x B(\theta_t, \phi'_x)) \ket{\Psi_{in}}
        \end{align}
        where $B(\theta_t,\phi'_x) \equiv \sum_{x}\left  [ B(\theta_t, \phi_x) \otimes \ket{x}\bra{x}\right ]$ at time $t$ and accelerating parameter $\theta_t$. The iterative form will be,
\begin{align} \label{iterate}
\begin{pmatrix}
\mathcal{A}_{x,t+1}\\
\mathcal{B}_{x,t+1}
\end{pmatrix} &= \begin{pmatrix}
\cos(\theta_t) & -i\sin(\theta_t) \\
0 & 0
\end{pmatrix} \begin{pmatrix}
\mathcal{A}_{x-1,t} \\
\mathcal{B}_{x-1,t}
\end{pmatrix} \nonumber \\
&+ \begin{pmatrix}
0 & 0\\
-ie^{i\phi_{x+1}}\sin(\theta_t) & e^{i\phi_{x+1}}\cos(\theta_t)
\end{pmatrix} \begin{pmatrix}
\mathcal{A}_{x+1,t} \\
\mathcal{B}_{x+1,t}
\end{pmatrix}.
\end{align} 
Fig.\,\ref{1D_disorder} shows the probability distribution of a single-particle accelerated quantum walk for different values of $a$ with spatial disorder in the phase parameter when $\theta_0 = \pi/2$.  In the plot we use the acronym ASQW - Accelerated Spatial-disordered Quantum Walk to identify the probability distribution due to spatial disorder in accelerated quantum walk. Particle delocalizes for higher value of $a$ implying that $\theta$ operator which saturates at maximum velocity $\theta (a,t) \rightarrow 0$ dominates over the random phase operator. Dispersion relations also gives us a picture of interplay between $\theta$ and $\phi$.

\noindent
{\it Temporal Disorder--}The temporal disorder in the single-particle quantum walk can be introduced by a time (step) dependent phase parameter $\phi_t$. $\phi_t$ can be randomly picked for each time-step from the range $0 \leq \phi_t \leq \pi$. The state after time $t$ with temporal disorder for accelerated quantum walk is given by,
        \begin{equation}
        \ket{\Psi(t)}_T = W_t(\theta_{t},\phi_t)...W_t(\theta_{2},\phi_2)W_t(\theta_1, \phi_1) \ket{\Psi_{in}}.
        \end{equation}
 The iterative form of the state of the particle at each position $x$ and time $(t+1)$ will be identical to Eq.(\ref{iterate}) with a replacement of $\phi_t$ in place of $\phi_x$. Fig.\,\ref{1D_disorder} shows the single-particle accelerated quantum walk for different values of $a$ with temporal disorder when $\theta_0 = \pi/2$. In the plot we use the acronym ATQW - Accelerated Temporal-disordered Quantum Walk. As the value of $a$ increases, the particle delocalizes in position space which implies that the interacting coin operator dominates over the phase operator. A similar behaviour as seen for spatial disordered case.

\subsection{Two-particle}

In order to introduce disorder in two-particle accelerated quantum walk, we are using the same methodology as used in single-particle accelerated quantum walk. That is, introduce a phase operator on each particle such that final operator on the particle will be interaction operator $C_{\theta_0}$ followed by phase operator $\Phi_{\phi}$. Phase operator for two-particle accelerated quantum walk will have the form,
\begin{equation}
\Phi_{\phi} = \begin{pmatrix}
1 & 0 \\
0 & \exp{i\phi}
\end{pmatrix} \otimes \begin{pmatrix}
1 & 0 \\
0 & \exp{i\phi}
\end{pmatrix}.   
\end{equation}\\
Therefore, the two-particle evolution operator with two parameters $\theta_0$ and $\phi$ before introducing disorder will have the form,
\begin{equation}
W(\theta_0, \phi) \equiv S_{(x,y)} (\Phi_{\phi}C_{\theta_0} \otimes \mathbb{I}_N)
\end{equation}
where,
\begin{widetext}
\begin{equation}
\Phi_\phi C_{\theta_0} = \begin{pmatrix}
\cos(\theta_0) & 0 & 0 & -i\sin(\theta_0) \\
0 & e^{i\phi}\cos(\theta_0) & -ie^{i\phi}\sin(\theta_0) & 0 \\
0 & -ie^{i\phi}\sin(\theta_0) & e^{i\phi}\cos(\theta_0) & 0 \\
-ie^{2i\phi}\sin(\theta_0) & 0 & 0 & e^{2i\phi}\cos(\theta_0)
\end{pmatrix}
\end{equation}
such that the dynamics equation when the initial state of the two particle walk is $|\Psi_{x_0,y_0}^2p(0)\rangle = |\psi_{c} \rangle \otimes|x_0\rangle \otimes |y_0\rangle$, is given by,
\begin{align} 
\Psi_{x,y_0}^{2p}(t+1) &= M^{+}_{(x-1,y_0)} \Psi_{x-1,y_0}^{2p}(t) + M^{-}_{(x+1,y_0)} \Psi_{x+1,y_0}^{2p}(t) \nonumber \\
\Psi_{x_0,y}^{2p}(t+1) &= M^{+}_{(x_0,y-1)} \Psi_{x_0,y-1}^{2p}(t) + M^{-}_{(x_0,y+1)} \Psi_{x_0,y+1}^{2p}(t) 
\label{eqDE}
\end{align} 
where, 
\begin{align}
M^{+}_{(x,y)} &= \begin{pmatrix}
\cos(\theta_{(x,y)}) & 0 & 0 & -i\sin(\theta_{(x,y)})\\
0 & 0 & 0 & 0 \\
0 & -ie^{i\phi_{(x,y)}}\sin(\theta_{(x,y)}) & e^{i\phi_{(x,y)}}\cos(\theta_{(x,y)}) & 0 \\
0 & 0 & 0 & 0
\end{pmatrix} \nonumber \\
 M^{-}_{(x,y)} &= \begin{pmatrix}
0 & 0 & 0 & 0 \\
0 & e^{i\phi_{(x,y)}}\cos(\theta_{(x,y)}) & -ie^{i\phi_{(x,y)}}\sin(\theta_{(x,y)}) & 0 \\
0 & 0 & 0 & 0 \\
-ie^{2i\phi_{(x,y)}}\sin(\theta_{(x,y)}) & 0 & 0 & e^{2i\phi_{(x,y)}}\cos(\theta_{(x,y)})
\end{pmatrix}. \nonumber
\end{align}
\end{widetext}
In the above expression parameters are position dependent such as, $\theta_0$ has been written as $\theta_{(x,y)}$ to represent the position dependent value in generic form. The final state at $(t+1)$ is $|\Psi^{2p}(t)\rangle = \sum_{x,y} (| \Psi_{x,y_0}^{2p}(t+1) \rangle + | \Psi_{x_0,y}^{2p}(t+1) \rangle)$. In Fourier-mode wave like solution form $\psi_{(x,y)}(t) = e^{-i\omega t} \psi_{x,y}$ and $\psi_{(x,y)} = e^{i(\kappa_1 x+ \kappa_2 y)} \psi(\kappa_1, \kappa_2)$ and $\Psi_{x,y_0}^{2p}= (\psi_{\uparrow\uparrow,(x,y_0)}; \psi_{\downarrow\downarrow,(x,y_0)}) = (\mathcal{A}_{x}; \mathcal{B}_{x})$ and $ \Psi_{x_0,y}^{2p}= (\psi_{\uparrow\downarrow,(x_0,y)}; \psi_{\downarrow\uparrow,(x_0,y)}) = ( \mathcal{C}_{y}; \mathcal{D}_{y})$ then,
\begin{align} 
e^{-i\omega}\mathcal{A}_{x} &= \cos(\theta_{x-1,y_0}) \mathcal{A}_{x-1} \nonumber \\
& - i\sin(\theta_{x-1,y_0}) \mathcal{B}_{x-1} \label{eq1} \\
e^{-i\omega}\mathcal{C}_{y} &= \cos(\theta_{x_0,y+1}) e^{i\phi_{x_0,y+1}} \mathcal{C}_{y+1} \nonumber \\
& - i\sin(\theta_{x_0,y+1}) e^{i\phi_{x_0,y+1}} \mathcal{D}_{y+1} \label{eq2}\\
e^{-i\omega}\mathcal{D}_{y} &= - i\sin(\theta_{x_0,y-1}) e^{i\phi_{x_0,y-1}} \mathcal{C}_{y-1} \nonumber \\
& + \cos(\theta_{x_0,y-1}) e^{i\phi_{x_0,y-1}} \mathcal{D}_{y-1} \label{eq3} \\
e^{-i\omega}\mathcal{B}_{x} &= - i\sin(\theta_{x+1,y_0}) e^{2i\phi_{x+1,y_0}} \mathcal{A}_{x+1} \nonumber \\
& + \cos(\theta_{x+1,y_0}) e^{2i\phi_{x+1,y_0}} \mathcal{B}_{x+1}. \label{eq4}
\end{align} 
From the Eqs.(\ref{eq1}- \ref{eq4}), it is clear that given coin operator  $\psi_{\uparrow\uparrow}$  and $\psi_{\downarrow\downarrow}$ evolves in superposition of $\psi_{\uparrow\uparrow}$ and $\psi_{\downarrow\downarrow}$ and similarly, $\psi_{\uparrow\downarrow}$  and $\psi_{\downarrow\uparrow}$ in superposition of $\psi_{\uparrow\downarrow}$ and $\psi_{\downarrow\uparrow}$, respectively as it was discussed for evolutions without phase operator. Therefore, the dispersion relation between the $\psi_{\uparrow\uparrow}$  and $\psi_{\downarrow\downarrow}$ for the given evolution operator when coin parameter at all sites is $ \theta_0$ and $\phi$ for two-particle accelerated discrete-time quantum walk is given by,
\begin{equation}
\cos(\omega + \phi) = \cos(\theta_0)\cos(\phi + \kappa_1).
\end{equation}
Similarly, the dispersion relation between $\psi_{\uparrow\downarrow}$  and $\psi_{\downarrow\uparrow}$  for two-particle when coin parameter at all sites is $\theta_0$ and $\phi$ is given by,
\begin{equation}
\cos(\phi + \omega) = \cos(\theta_0) \cos(\kappa_2).
\end{equation}

The expression for state of two-particle discrete-time quantum walk shows that coin state $|\uparrow\uparrow\rangle $ and $| \downarrow \downarrow\rangle$ will evolve the particles in $x-$direction and  coin state $|\uparrow\downarrow\rangle $ and $| \downarrow \uparrow\rangle$ will evolve the particles in $y-$direction in two-dimension when the initial state of the particle is symmetric or same. Therefore, above formulation can be written in one-dimensional position space when the initial state in coin space is one of the four basis state. In two dimensional position space, using transfer matrix with the redefined component fields, $\Psi_{(x,y_1)}^{2p} = (\psi_{\uparrow\uparrow,(x,y_1)}, \psi_{\downarrow\downarrow,(x-1,y_1)}) = (\mathcal{A}_{x}; \mathcal{B}_{x-1})$ and $\Psi_{(x_1,y)}^{2p} = (\psi_{\uparrow\downarrow,(x_1,y)}, \psi_{\downarrow\uparrow,(x_1,y-1)}) = (\mathcal{C}_{y}; \mathcal{D}_{y-1})$ is given by,
\begin{align} 
\Psi_{x+1,y_1}^{2p} &= T_{x,y_1}\Psi_{x,y_1}^{2p} \nonumber \\
\Psi_{x_1,y+1}^{2p} &= T_{x_1,y}\Psi_{x_1,y}^{2p}.
\label{spatial}
\end{align}
The transfer matrix $T_{x,y}$ at position $(x,y)$ will be of the form,
\begin{widetext}
\begin{equation}
T_{x,y} = \begin{pmatrix}
e^{i\omega} \sec(\theta_{x,y}) & 0 & 0 & -ie^{-2i\phi_{x,y}}\tan(\theta_{x,y}) \\
0 & e^{-i(\omega + \phi_{x,y})}\sec(\theta_{x,y}) & i\tan(\theta_{x,y}) & 0 \\
0 & -i\tan(\theta_{x,y}) & e^{i(\omega + \phi_{x,y})} \sec(\theta_{x,y}) & 0  \\
i\tan(\theta_{x,y}) & 0 & 0 & e^{-i(\omega + 2\phi_{x,y})}\sec(\theta_{x,y})
\end{pmatrix}. \label{eq5}
\end{equation} 
\end{widetext}
Eqs. \eqref{spatial} and \eqref{eqDE} shows that introducing disorder in either of the coin parameter $\theta_x$ or $\phi_x$ will lead to spatial and temporal disorder, respectively. Therefore, like it was done for single-particle quantum walk we will introduce disorder in $\phi_x$ and we will accelerate parameter $\theta$ as, $\theta_t = \theta_0 e^{-at}$ where $a$ is the acceleration in the two-particle accelerated walk. For further study of two particle quantum walk, we will analyse the initial state $|\uparrow \uparrow \rangle \otimes \ket{x=0} \otimes \ket{y=0} $ therefore the study reduces to one-dimension as evolution in $H_{p_y}$ is zero for the given initial state.
\begin{figure}[h!]
\centering 
\includegraphics[width = \linewidth]{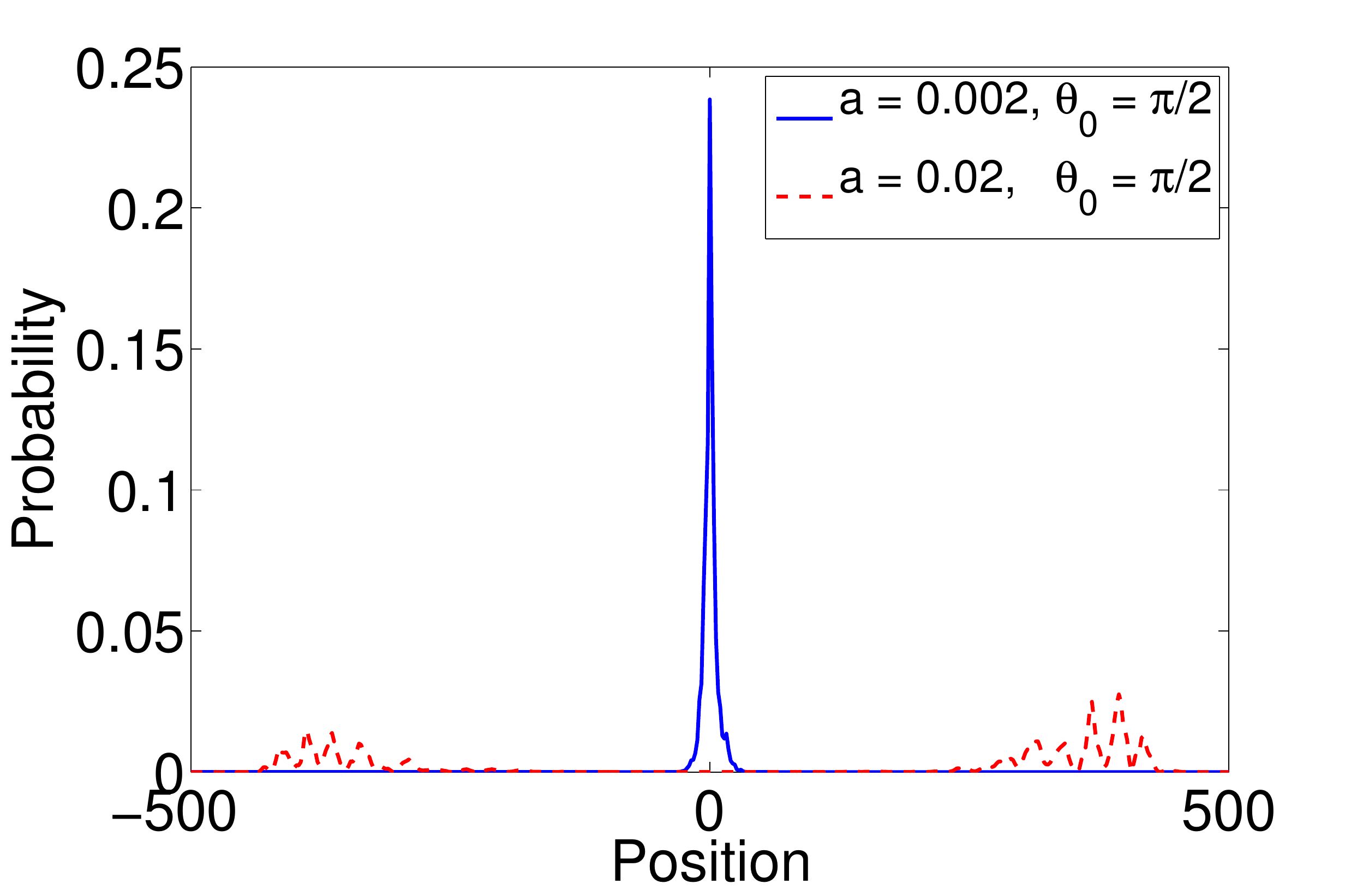}
\caption{Probability distribution for two-particle accelerated quantum walk with spatial disorder  for different value of $a$ within Hilbert-space $\mathcal{H}_{p_x}$. Coin operator is an entangling operator followed by the phase operator and initial state is $\ket{\psi_{in}} = \ket{\uparrow\uparrow} \otimes \ket{x = 0} \otimes \ket{y=0}$. For this initial state probability distribution in  Hilbert-space $\mathcal{H}_{p_y}$ is zero. The particle delocalizes for higher value of $a$.}
\label{Prob_SQW}
\end{figure}
\begin{figure}[h!]
\centering
\includegraphics[width = \linewidth]{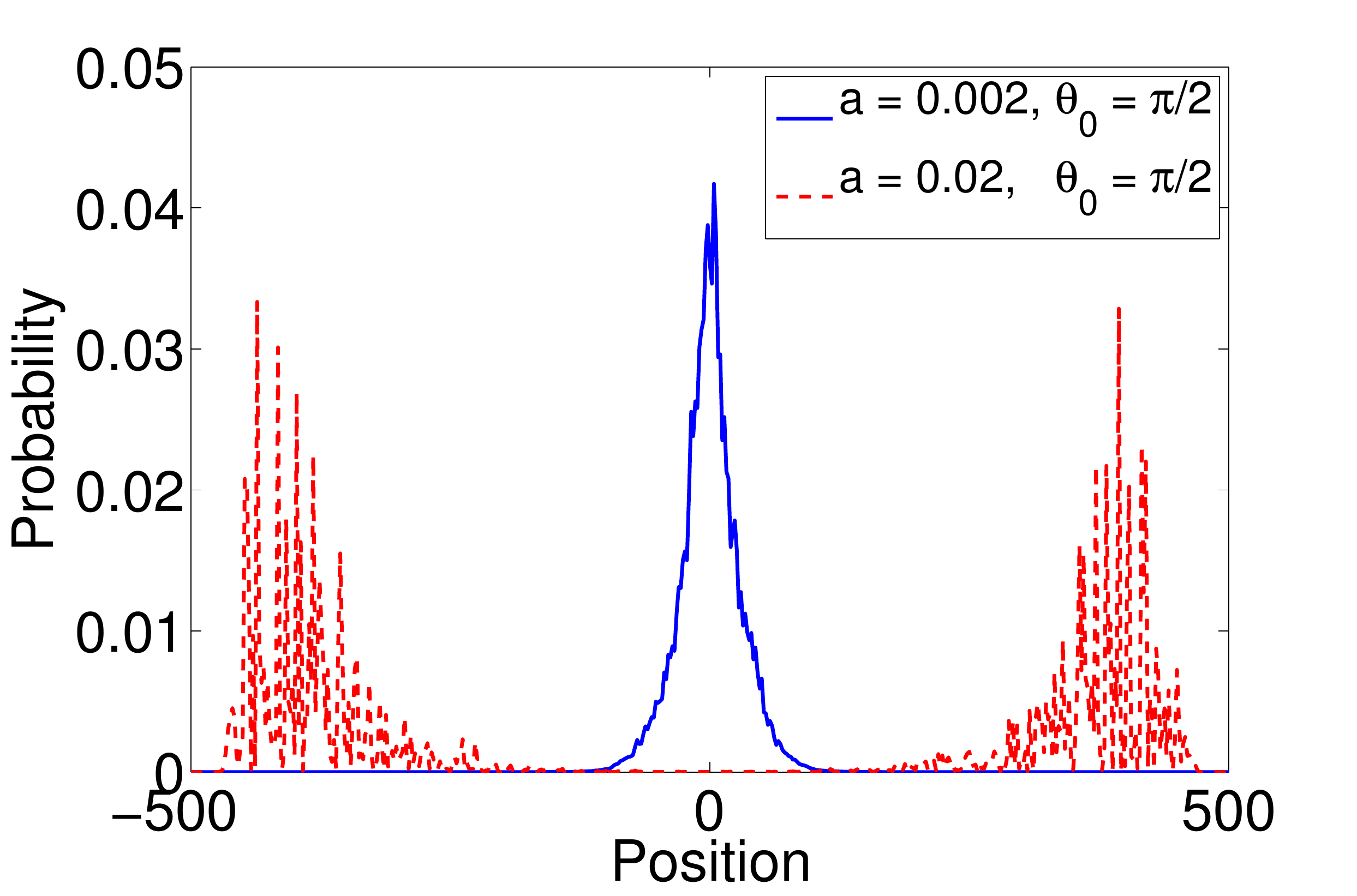}
\caption{Probability distribution two-particle accelerated quantum walk with temporal disorder for different value of $a$ within Hilbert-space $\mathcal{H}_{p_x}$. Two-particle coin operation is followed by the phase operator and initial state is $\ket{\psi_{in}} = \ket{\uparrow\uparrow} \otimes \ket{x = 0} \otimes \ket{y=0}$. For this initial state probability distribution in  Hilbert-space $\mathcal{H}_{p_y}$ is zero. The particle delocalizes for higher value of $a$.}
\label{Prob_TQW}
\end{figure}



\noindent
{\it Spatial Disorder--}The spatial disorder in two-particle accelerated quantum walk in one dimension is introduced by a position dependent phase parameter $\phi_x$ as from the transfer matrix in Eq. \eqref{eq5} it can be seen that $\Psi_x$ depends on the phase parameter $\phi_x$ moving from one site to another. $\phi_x$ can be randomly picked for each position from the range $0 \leq \phi_x \leq \pi$. The state after time $t$ with spatial disorder using position dependent phase parameter $\phi_x$ and accelerated $\theta$ parameter will be,
       \begin{equation}
        \ket{\Psi_{x}^{2p}(t)}_S = W_x(\theta_{t},\phi'_x)...W_x(\theta_{2},\phi'_x)W_x(\theta_1, \phi'_x) \ket{\Psi_{in}^{2p}},
        \end{equation}
where $W(\phi'_x) \equiv \sum_{x}\left[W(\phi_x) \otimes \ket{x}\bra{x}\right]$. The iterative form of the state of the particle at each position $x$ and time $(t+1)$ will be identical to Eq. \eqref{eqDE} by replacing $\phi$ with $\phi_x$. Fig.\ref{Prob_SQW} shows the accelerated two-particle quantum walk for different values of $a$ with spatial disorder. Higher the value of $a$, the particle delocalizes which implies that at higher value of $a$ the interaction operator $C_{\theta}$ with $\theta = \theta_{0} e^{-at}$ dominates over the phase operator in coin operation. Fig. \ref{Prob_SQW} shows probability distribution for different value of acceleration. As the acceleration in the system increases, the particles in position space delocalizes as the entangling operator dominates over the phase operator.

\begin{figure}[h!]
\centering
\includegraphics[width = \linewidth]{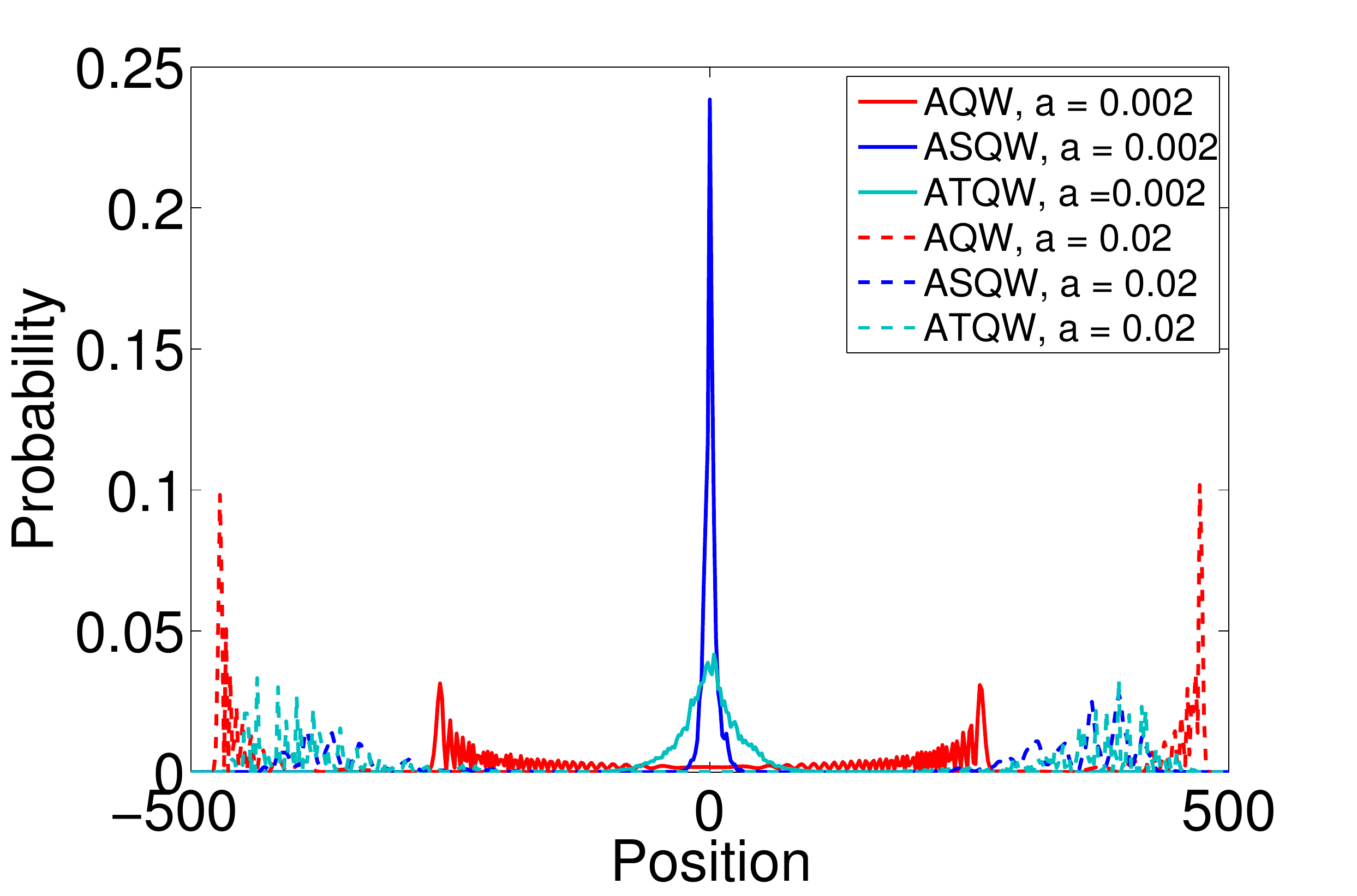}
\caption{Probability distribution for two-particle accelerated quantum walk (AQW) in position Hilbert space $\mathcal{H}_{p_x}$ with spatial (ASQW) and temporal (ATQW) disorder for different value of $a$. Coin operator is an entangling operator followed by the phase operator and initial state is $\ket{\psi_{in}} = \ket{\uparrow\uparrow} \otimes \ket{x = 0} \otimes \ket{y=0}$. The particle delocalizes for higher value of $a$. The spread is always maximum even when compared to the delocalized distribution with disorder.}
\label{Prob}
\end{figure}

\noindent
{\it Temporal Disorder--}The temporal disorder in the two-particle discrete-time quantum walk can be introduced by a step dependent phase parameter $\phi_t$,  randomly picked for each step from the range $0 \leq \phi_t \leq \pi$. The state after time $t$ with temporal disorder using step dependent phase parameter $\phi_t$ and accelerated $\theta$ parameter will be,
     \begin{equation}
       \ket{\Psi^{2p}(t)}_T = W_t(\theta_{t},\phi_t)...W_2(\theta_{2},\phi_2)W_1(\theta_1, \phi_1) \ket{\Psi_{in}^{2p}}.
        \end{equation}       
The iterative form of the state of the particle at each position $x$ and time $(t+1)$ will be identical to Eq.\,(\ref{eqDE}) with only a replacement of $\phi_{(x,y)}$ with $\phi_t$. Fig.\,\ref{Prob_TQW} show the probability distribution of the  accelerated two-particle quantum walk for different values of $a$ with temporal disorder. Higher the value of $a$, the particle delocalizes which implies that at higher value of $a$ the interaction coin operator dominates over the phase operator which is similar to the spatial disordered and single-particle case.  Fig.\,\ref{Prob} shows a comparison of the probability distribution for accelerated two-particle quantum walk,  with and without spatial and temporal disorder for different value of $a$ when $\theta_0 = \pi/2$. Spatial disorder for small value of $a$ shows a strong localization compared to the temporal disorder.  When acceleration dominates over disorder the probability distribution delocalizes for both, spatial and temporal disorder. This can be seen by comparing the distribution for $a=0.02$ (delocalized) and $a=0.002$ (localized). However, spread of the distribution is always wider in absence of disorder when compared configuration of leading to deloclization in presence of disorder. 
\begin{figure}[h!]
\centering
\includegraphics[width = \linewidth]{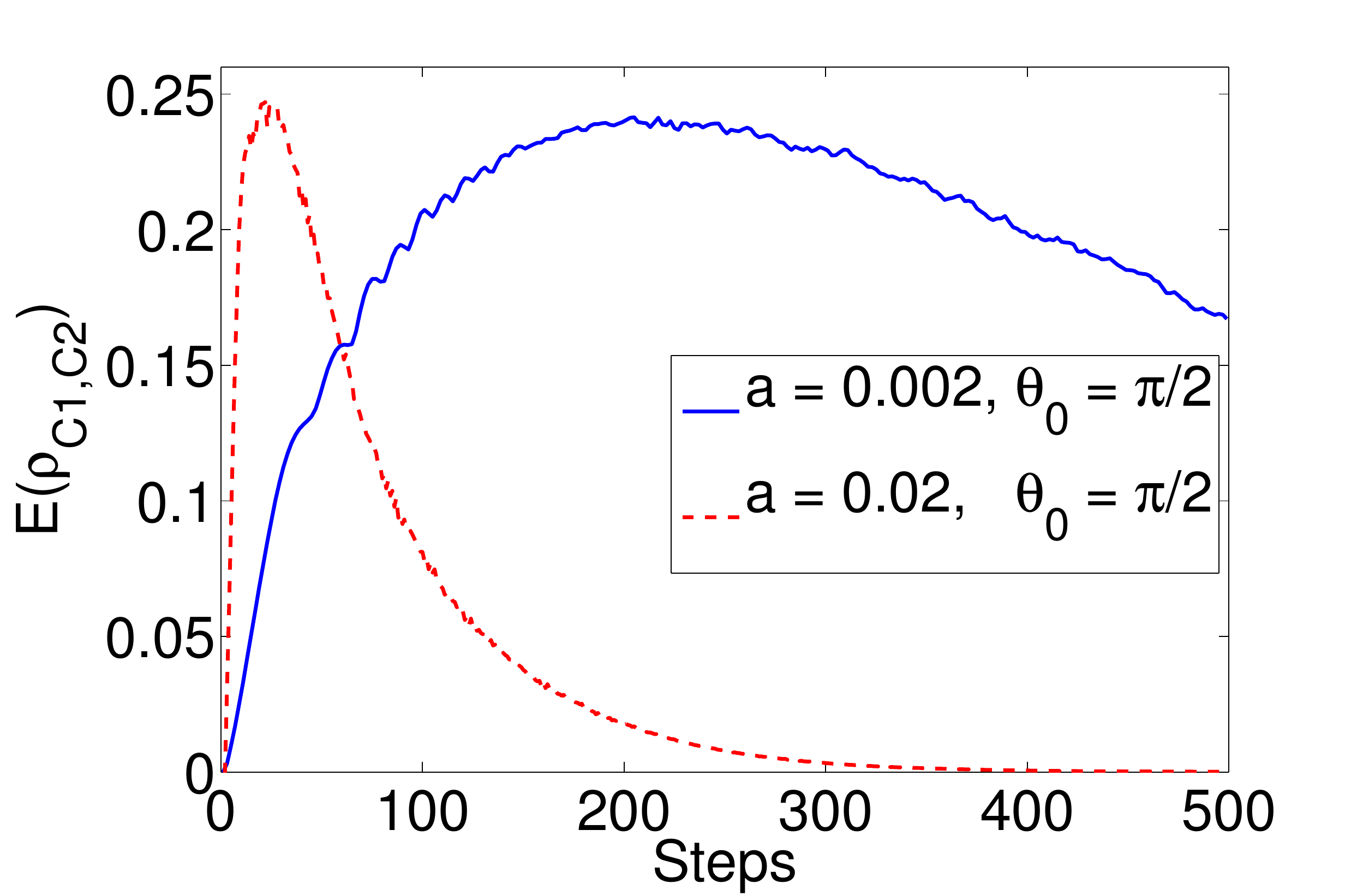}
\caption{Entanglement between the particles of coin Hilbert space in two-particle accelerated quantum walk with spatial disorder (ASQW) for different value of $a$. Initial state is $\ket{\psi_{in}} = \ket{\uparrow\uparrow} \otimes \ket{x = 0} \ket{y=0}$ averaged over 1000- runs. For larger $a$ entanglement dies faster in time. }
\label{Nega_SQW}
\end{figure}
\begin{figure}[h!]
\centering
\includegraphics[width = \linewidth]{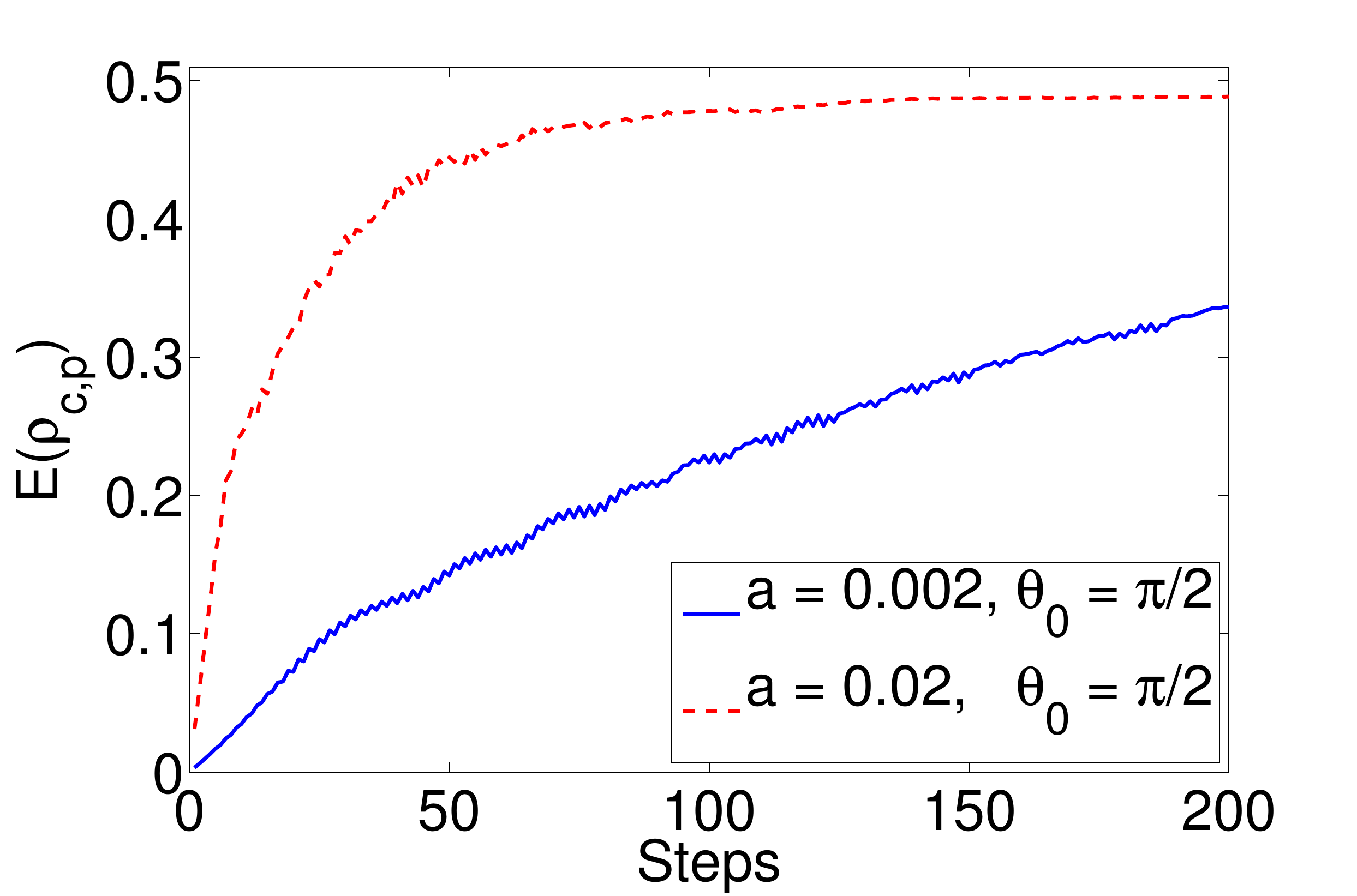}
\caption{Entanglement between the coin and position Hilbert space $\mathcal{H}_{p_x}$ in two-particle accelerated quantum walk with spatial disorder (ASQW) for different value of $a$ as function of time (steps). Initial state is $\ket{\psi_{in}} = \ket{\uparrow\uparrow} \otimes \ket{x = 0} \otimes \ket{y=0}$ averaged over 500- runs. Entanglement between coin and position Hilbert space $\mathcal{H}_{p_y}$ is zero as the probability distribution in this Hilbert space is zero for the given initial state.  For larger $a$ entanglement reaches maximum value faster in time.}
\label{Entangle_SQW}
\end{figure}
\begin{figure}[h!]
\centering
\includegraphics[width = \linewidth]{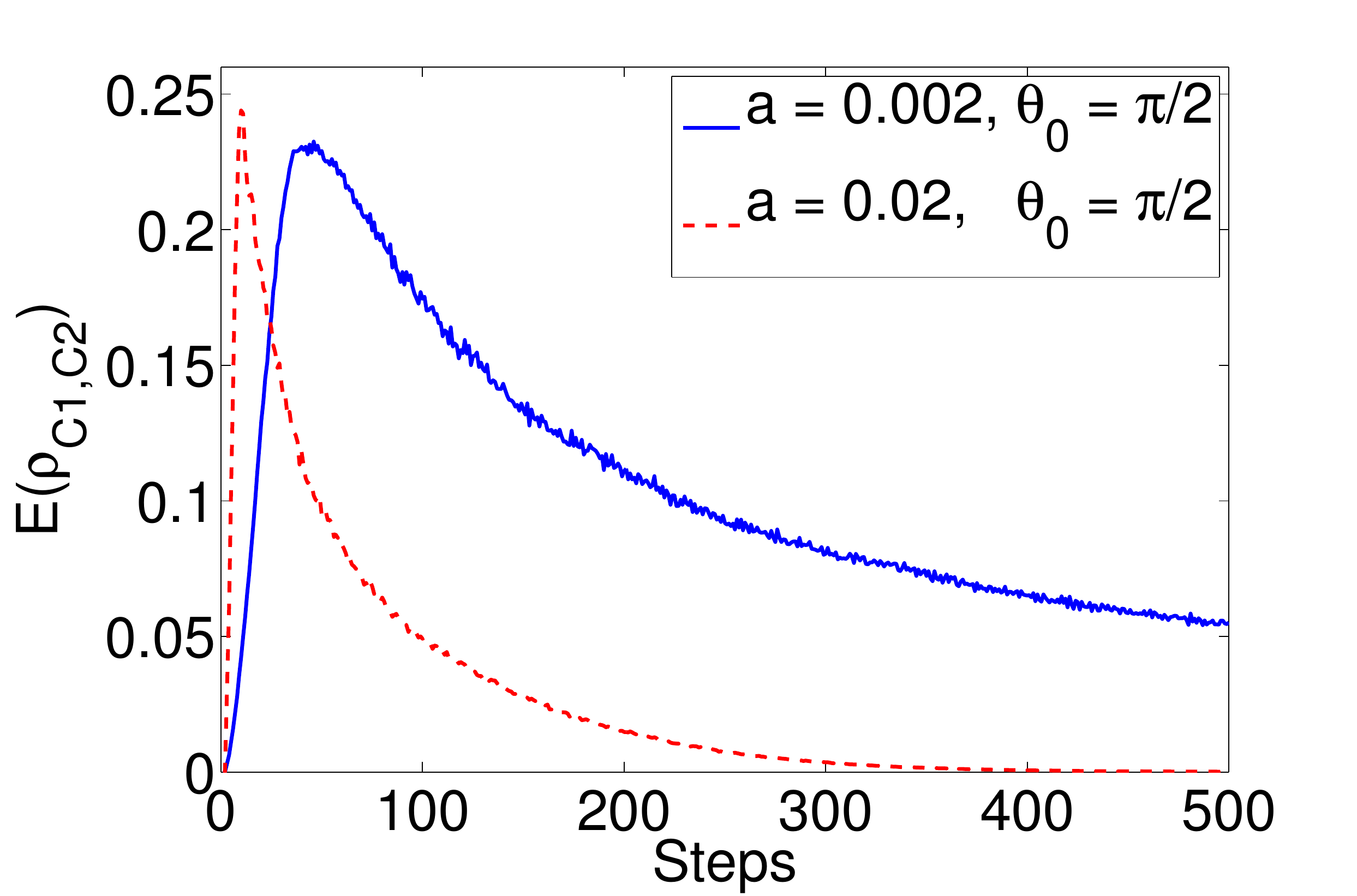}
\caption{Entanglement between the particles of coin Hilbert space in two-particle accelerated quantum walk with temporal disorder (ATQW) for different value of $a$. Initial state is $\ket{\psi_{in}} = \ket{\uparrow\uparrow} \otimes \ket{x = 0} \otimes \ket{y=0}$ averaged over 1000- runs. For larger $a$ entanglement dies faster in time.}
\label{Nega_TQW}
\end{figure}
\begin{figure}[h!]
\centering
\includegraphics[width = \linewidth]{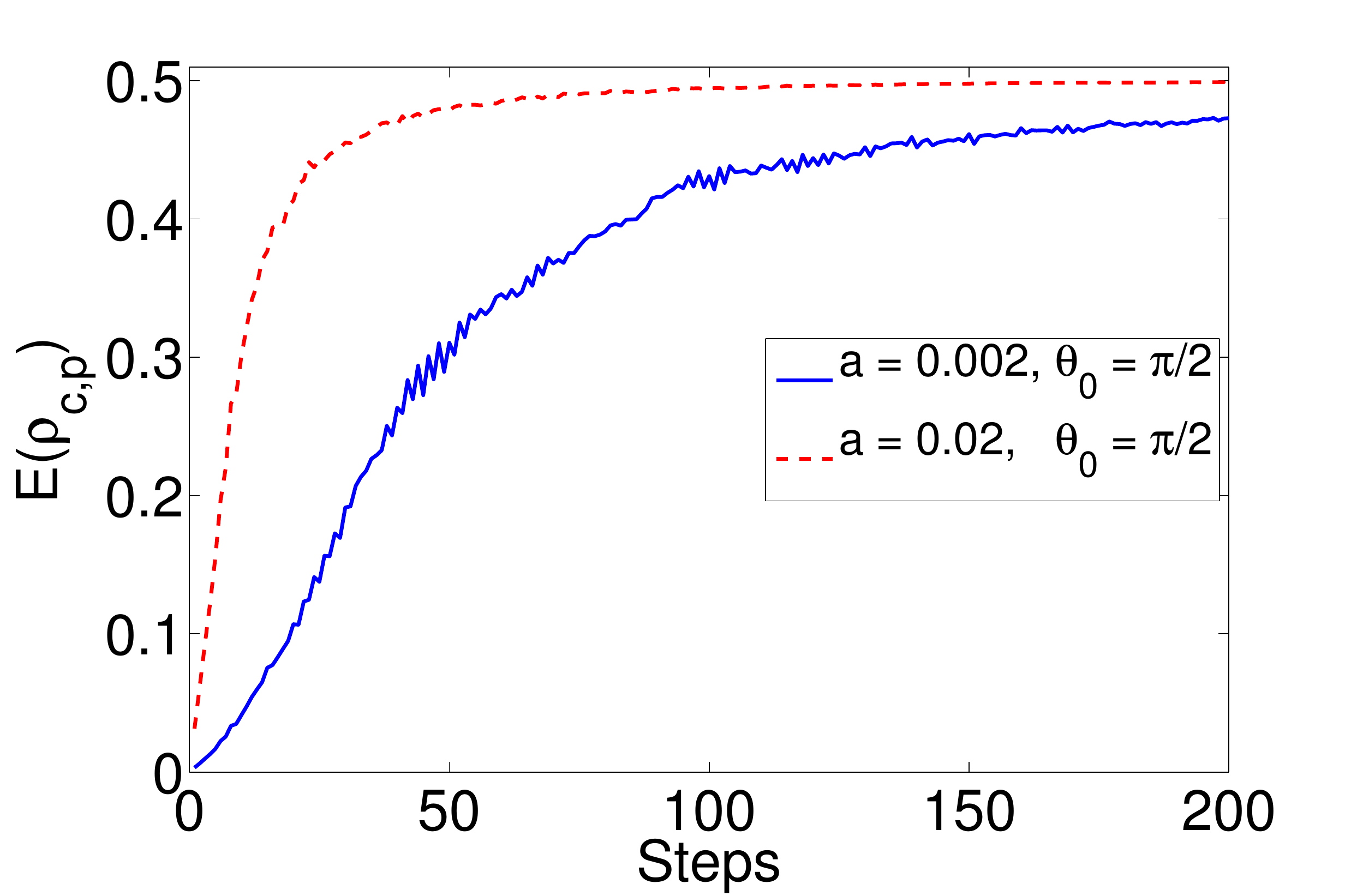}
\caption{Entanglement between the coin and position Hilbert space $\mathcal{H}_{p_x}$ in two-particle accelerated quantum walk with temporal disorder (ATQW) as function of time (steps) for different value of $a$. Initial state is $\ket{\psi_{in}} = \ket{\uparrow\uparrow} \otimes \ket{x = 0} \otimes \ket{y=0}$ averaged over 500- runs. Entanglement between coin and position Hilbert space $\mathcal{H}_{p_y}$ is zero for the given initial state. For larger $a$ entanglement reaches maximum value faster in time.}
\label{Entangle_TQW}
\end{figure}
 \begin{figure}[h!]
\centering
\includegraphics[width = \linewidth]{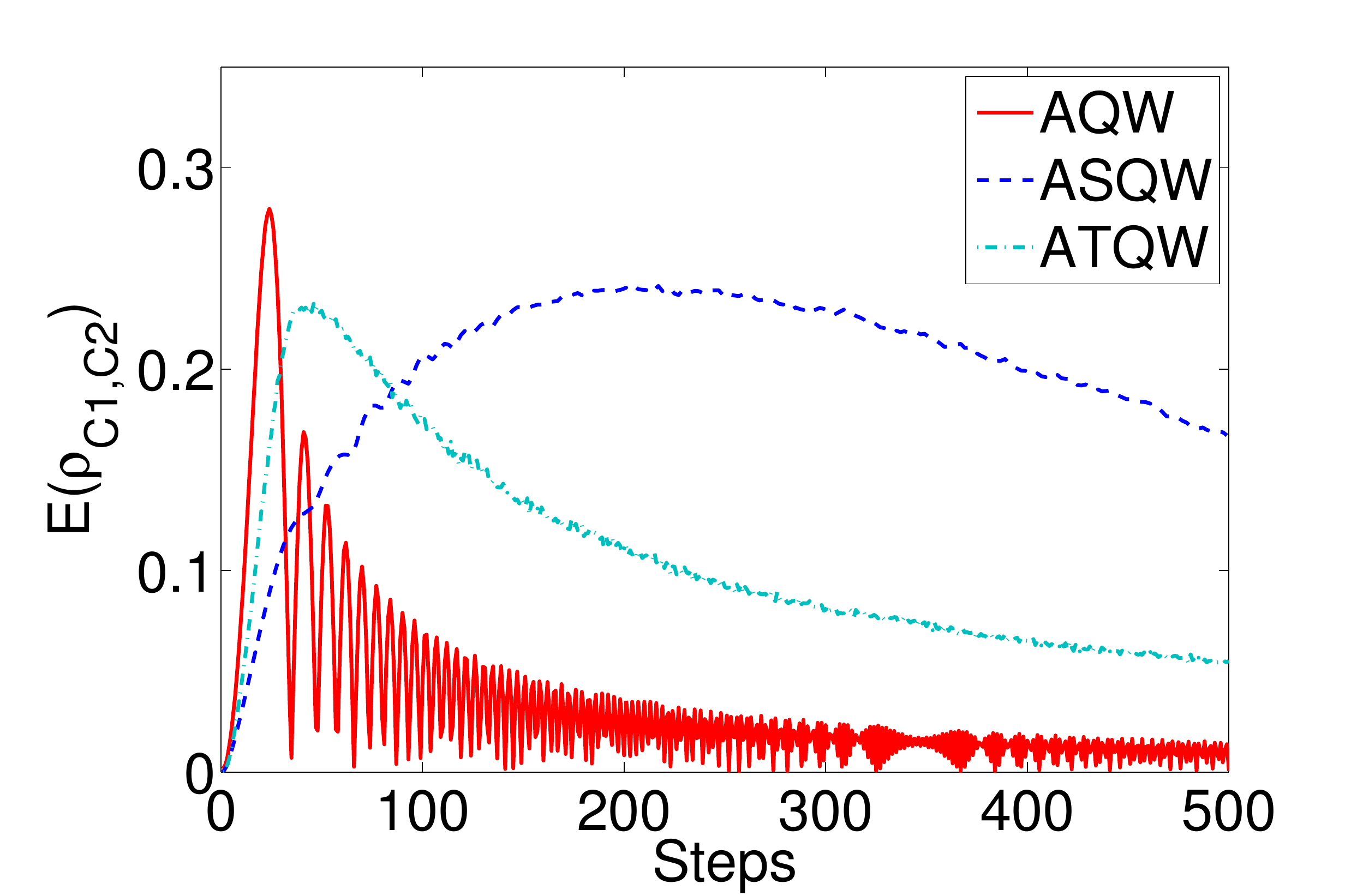}
\caption{Comparison of entanglement between the coin Hilbert spaces for two-particle accelerated quantum walk with $a = 0.002$ which is the localized case. Two-particle coin operator with $\theta_0 = \pi/2$ is followed by the random phase operator and initial state is $\ket{\psi_{in}} = \ket{\uparrow\uparrow} \otimes \ket{x = 0} \otimes \ket{y=0}$ averaged over 1000- runs.  Entanglement dies of faster for accelerated quantum walk when compared to accelerated quantum walk with disorder which stay higher for significantly longer time.}
\label{Nega_0_002}
\end{figure}
\begin{figure}[h!]
\centering
\includegraphics[width = \linewidth]{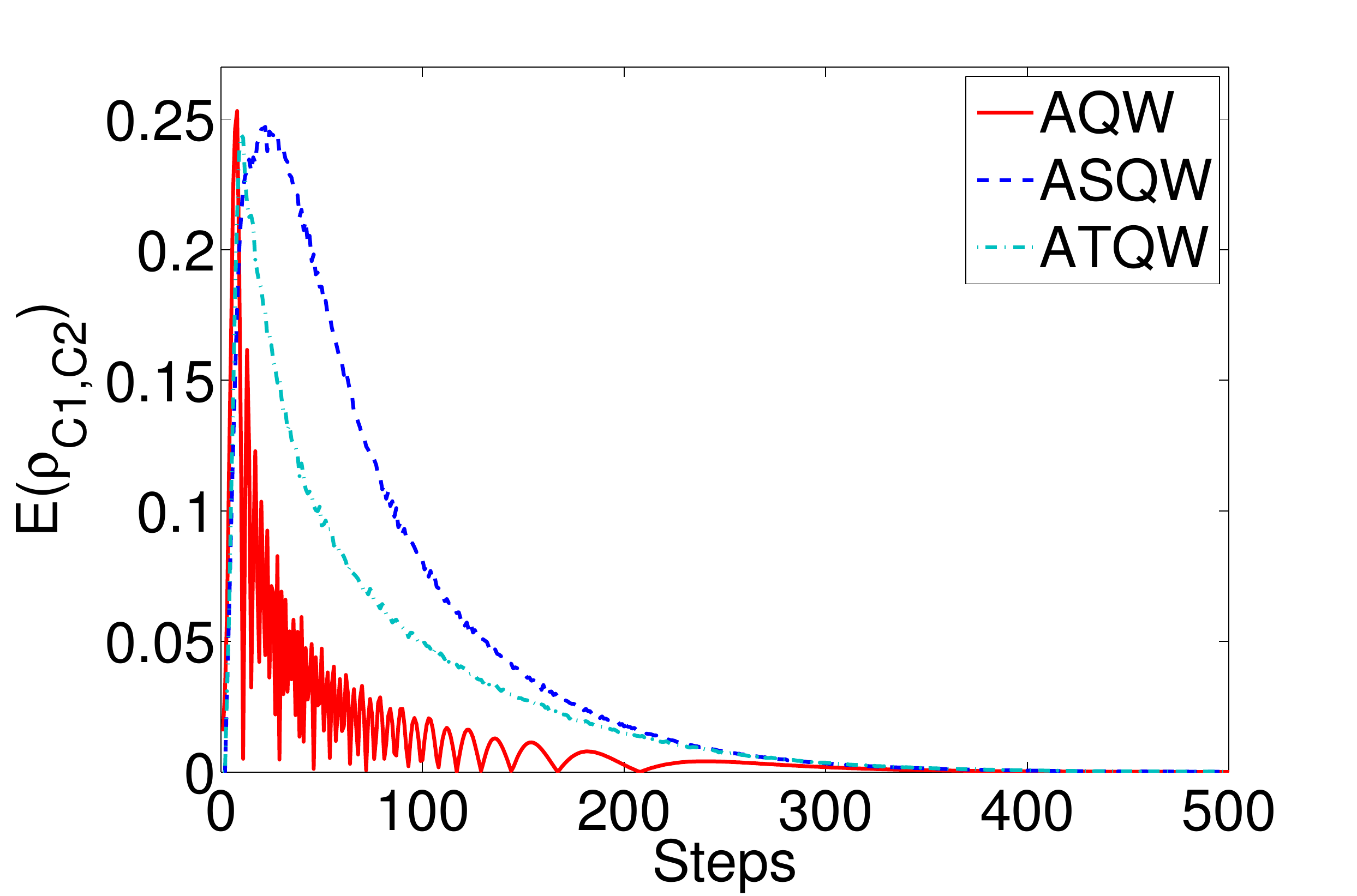}
\caption{Comparison of entanglement between the coin Hilbert spaces for two-particle accelerated quantum walk with $a = 0.02$ which is the delocalized case. Two-particle coin operation $\theta_0 = \pi/2$  is followed by the random phase operator for spatial and temporal disordered accelerated quantum walks. The initial state is $\ket{\psi_{in}} = \ket{\uparrow\uparrow} \otimes \ket{x = 0} \otimes \ket{y=0}$ averaged over 1000- runs. Entanglement dies of faster for accelerated quantum walk when compared to accelerated quantum walk with disorder.}
\label{Nega_0_02}
\end{figure}


{\it Entanglement in two-particle accelerated quantum walk with disorder - } Figs.\, \ref{Nega_SQW} and \,\ref{Nega_TQW} shows negativity as measure of entanglement between the two particles for different values of $a$ for a two-particle accelerated quantum walk with spatial disorder and temporal disorder, respectively. Figs.\,\ref{Entangle_SQW} and \,\ref{Entangle_TQW} shows negativity as measure of entanglement between the particle and position Hilbert space for different values of $a$ for a two-particle  accelerated quantum walk with spatial disorder and temporal disorder, respectively.
 
The entanglement for the delocalized case decays faster then the localized case in two-particle accelerated quantum walk with spatial and temporal disorder. This can be seen by comparing the values in the Figs.\,\ref{Nega_QW}, \,\ref{Nega_SQW} and \,\ref{Nega_TQW}, respectively. For $a=0.002$ when the particle is localized, entanglement decays faster but entanglement reaches to its maximum value slower in localized case when compared to delocalized case when $a=0.02$. Entanglement decays faster for two-particle accelerated quantum walk without disorder when compared to walk with spatial and temporal disordered system in one dimension as shown in Figs.\,\ref{Nega_0_002} and \,\ref{Nega_0_02}. It can be seen that maximum entanglement between the particle is found in the accelerated quantum walk but entanglement stays for longer time for spatial localized accelerated quantum walk. It happens because in localized case both the particle's probability amplitude superimpose in localized position space with stronger interference to give entanglement between the particles and hence its decays slower then the entanglement between the delocalized case. Maximum the spread in the position space faster is the death of the entanglement between the particles evolving in position space.  The entanglement between the particle and position space for standard accelerated quantum walk, accelerated quantum walk with spatial disorder and temporal disorder for the two-particle walk is shown in Figs.\, \ref{Entangle_QW}, \,\ref{Entangle_SQW} and \,\ref{Entangle_TQW}, respectively. Similar to single-particle accelerated quantum walk, for higher value of $a$ particle reaches the maximum entanglement in short time for standard accelerated quantum walk as well as for disordered quantum walk and remains constant with time.  

\section{\label{sec4} Discussion and concluding remarks}


In this work we have introduced accelerated discrete-time quantum walk and studied the dynamics of single-particle and two-particle system. For a single-particle quantum walk, introduction of acceleration into the system increases the rate of spread of the probability amplitude over the position space. It also results in enhancement of entanglement between the particle and the position space and reaches a maximum value in shorter time. Entanglement between the position and  particle in the single-particle quantum walk reaches its maximum value faster with increase in acceleration and then saturates at the maximum value. The increase in entanglement seen with time dependent coin operation causing acceleration allies well with the earlier results of increase in entanglement between particle and position space due to random coin operation for each time~\cite{CMC, VAR13}. Thus, one can corroborate that the enhancement of entanglement is in general due to the time dependent quantum coin operation and enhancement due to randomness in time is a specific case.  

In two-particle case, the dynamics was defined by taking into account the bosonic and fermionic nature into account and that restricted the dynamics to only one dimensional position space. For the two-particle case also the acceleration in the system spreads the probability amplitude faster in position space.  In a regular quantum walk the maximum spread is achieved only when $\theta_0 \rightarrow 0$ but that happen without any interference of the particle in position space. But in accelerated quantum walk same maximum spread is achieved along with interference effect. We have explicitly studied the effect of acceleration parameter $a$ and how the rate at which the probability distribution spread and entanglement of the system change.

In a standard quantum walk (without acceleration) with disorder we will see localization but in accelerated quantum walk with disorder localization is seen only when the acceleration is very small. As acceleration increases the system becomes delocalized because, the increase in acceleration increases the perturbation in the unitary operator and that dominates over the disordered phase parameter in the system. From this study we can see a promise in the direction of modelling the dynamics of quantum systems where the transition from localized to delocalized state is an important phenomena. 
For both, spatial and temporal disordered quantum walk with acceleration, the entanglement between the two particles increase with acceleration, reaches the maximum entanglement value faster and decays faster too in time. This implies that the entanglement between the interacting particles stays for a longer time when the value of acceleration is smaller when compared to higher value of acceleration for which the particle is delocalized. It could be because the  superposition and interference between the particles wave function in two-particle accelerated quantum walk is more than the interference effect in localized discrete-time quantum walk.


The two-particle quantum walk system in this paper is similar to the condensed matter Ising model with is a special case of Heisenberg model. The generalised Heisenberg model for $N$ spin-1/2 particle is described by the Hamiltonian,
\begin{widetext}
\begin{equation}
H_{\gamma} = \sum_{i}^{N-1}\left[ (1+\gamma)(S^{x}_{i} \otimes S^{x}_{i+1}) + (1-\gamma)\lbrace(S^{y}_{i} \otimes S^{y}_{i+1}) + (S^{z}_{i} \otimes S^{z}_{i+1})\rbrace\right]
\end{equation} 
\end{widetext}
where $S^{\alpha} = \sigma_{\alpha}$ with $\alpha = x, y, z$ are Pauli matrices and $\gamma$ is the degree of anisotropy in the system. As $\gamma \rightarrow 1$ model tends to the Ising model with the spin components in the $x$- direction completely ordered and $y-$ and $z-$ component completely disordered. We have considered a one-dimensional two-particle discrete time quantum walk to be ordered in x- direction therefore $\gamma = 1$ and since its a two-particle interaction therefore $N=2$ and hence the Hamiltonian for two-particle quantum walk is given by,
 \begin{equation}
 H = S^{x}_1 \otimes S^{x}_2,
 \end{equation}
where $S^{x} = \sigma_x$ which is Pauli matrix.
The accelerated two-particle quantum walk is similar to the Ising model with perturbed Ising model. The acceleration is introduced in the system by accelerating the quantum coin angle. The coin operation is given by,
\begin{equation}
C_{\theta_t} = \exp^{-i\theta_t H}
\end{equation} 
where $\theta_t = \dfrac{\pi}{2} \exp^{-at}$ and $a$ is the acceleration in the system.
Disorder in the system is introduced by randomizing the value of $\phi$ in the Phase operator. randomizing $\phi$ at each position in the position space gives Anderson localization and randomizing $\phi$ at each step in the position space gives weak localization.

The connection of quantum walks dynamics with Dirac Hamiltonian and Dirac cellular automata along with this intriguing connections between acceleration, entanglement generation and localization paves way for further investigation towards understanding the role of acceleration, mass and entanglement in relativistic quantum mechanics and quantum field theory.
This work shows a promise in the direction of modelling accelerated single-particle and accelerated - interacting two-particle quantum system dynamics using controllable quantum walks. Though these studies provide an additional operational tools for quantum simulations using quantum walks, more analytical studies are still required to explicitly formalize some of the observations related to entanglement behaviour in the system with more than two particle. 

\vskip 0.2in
\noindent
{\bf Acknowledgment:}\\
\\
\noindent
CMC would like to thank Department of Science and Technology, Government of India for the Ramanujan Fellowship grant No.:SB/S2/RJN-192/2014. CMC and RL would also acknowledge the support from US Army Research Laboratory.

\end{document}